\documentclass[a4paper,fleqn,usenatbib]{mnras} 

\usepackage{newtxtext,newtxmath}

\usepackage[T1]{fontenc}
\usepackage{ae,aecompl}

\usepackage{graphicx}	
\usepackage{amsmath}	
\usepackage{amssymb}	

\title[Orbital period modulation in hot Jupiter systems]{Orbital period modulation in hot Jupiter systems}

\author[A. F. Lanza]{
A. F. Lanza,\thanks{E-mail: antonino.lanza@inaf.it}
\\
INAF-Osservatorio Astrofisico di Catania, Via S.~Sofia,78 - 95123 Catania, Italy\\
}

\date{Accepted XXX. Received YYY; in original form ZZZ}

\pubyear{2020}

\begin{document}
\label{firstpage}
\pagerange{\pageref{firstpage}--\pageref{lastpage}}
\maketitle

\begin{abstract}
We introduce a model for the orbital period modulation in systems with close-by giant planets based on a spin-orbit coupling  that transfers angular momentum from the orbit to the rotation of the planet and viceversa. The coupling is produced by a permanent non-axisymmetric gravitational quadrupole moment assumed to be present in the solid core of the planet. We investigate two regimes of internal planetary rotation, that is, when the planet rotates rigidly and when {the rotation of its deep interior is time dependent as a consequence of a vacillating or intermittent convection in its outer shell.} The model is applied to a sample of very hot Jupiters predicting maximum transit-time deviations from a constant-period ephemeris of approximately 50~s in the case of rigid rotation. The transit time variations of WASP-12, currently the only system showing evidence of a non-constant period, cannot be explained by assuming rigid rotation, but can be modelled in the {time-dependent internal rotation regime}, thus providing an alternative to their interpretation in terms of a tidal decay of the planet orbit. 
\end{abstract}




\section{Introduction}
\label{intro}
Hot Jupiters (HJs) are giant planets orbiting closer than $\sim 0.15$~au to their host stars. Transiting systems with orbital periods shorter than $\sim 1.5$~days have been the subject of long-term timing observations with the purpose of detecting the expected tidal decay of their orbits. Specifically, tides extract angular momentum from the orbital motion of the HJs to spin up their host stars the rotation period of which is generally longer than  the orbital period. In almost all the cases, the total angular momentum of the system is insufficient to reach a stable equilibrium with the stellar rotation synchronized to the planet orbit \citep{Levrardetal09,DamianiLanza15}, thus the ultimate fate of most of the HJs is to experience a final orbital decay and transfer mass to their host stars via Roche lobe overflow \cite[e.g.][]{Valsecchietal14}. 

The observational signature of a tidal orbital decay is a decrease of the orbital period. In the case of transiting HJs, this can be measured through the time difference between the observed ($O$) and predicted ($C$) epochs of mid transits on the basis of a constant-period ephemeris,  $O-C$ becoming increasingly negative as time passes by. In the case of a constant period derivative, $|O-C|$  increases with the square of the number of elapsed orbital periods since a reference epoch.  A sample of systems particularly suitable to measure the expected tidal orbital decay has been recently discussed by \citet{Patraetal20}. 

In view of the interest in measuring the tidal orbital decay of HJs, it is worth investigating phenomena that could lead to variations of the epochs of mid-transits that could be misinterpreted as evidence for the searched decay. These include the precession of the line of the apsides in the case of slightly eccentric orbits, the light-time effect induced by a distant third body in the system \citep[e.g.,][]{Boumaetal20}, and processes occurring in the interior of the host stars  producing a variation of their gravitational quadrupole moment \citep{Applegate92,Lanzaetal98,WatsonMarsh10} or a spin-orbit coupling with the planetary orbit \citep{Lanza20}. 

All these mechanisms produce an oscillation of the $O-C$'s, but, since the oscillation period is of the order of several decades, they can be misinterpreted as a tidal orbital decay when the observations are not extended enough to reveal  the change in sign of the period derivative. 

In the present paper, we introduce another mechanism that can produce a long-term modulation of the $O-C$'s in the case of very close-by hot Jupiters and discuss its possible role, focusing on the case of WASP-12, currently the most promising system for an observational detection of orbital period changes, with a planet  of $\sim 1.5$ Jupiter masses on a 1.094-day orbit around an F-type main-sequence star.   

\section{Observations}
Among the systems recently considered by \cite{Patraetal20}, only WASP-12 shows a significant deviation of its $O-C$ curve from a linear trend, that can be interpreted as an orbital period decay. For the next best case, WASP-19, current evidence is only marginal and further observations are required to confirm it. 

Alternative models to interpret the $O-C$ diagram of WASP-12 have been proposed, but they seem to be less likely than  a decrease of the orbital period. \cite{Yeeetal20} discussed the case of a slightly eccentric orbit with a precession of the line of the apsides; the presence of a third body in the system; and the possible effect of mass transfer finding all of them less likely than the tidal decay interpretation. Apsidal motion would produce oppositely oscillating $O-C$'s for the transits and planet occultations (secondary eclipses)  that should already be observable after a decade of nearly continuous monitoring, while a third body would induce a long-term acceleration of the barycentre of the system that should be detectable in the current long-term series of radial-velocity measurements. Finally, mass exchange between the planet and the star, that is likely to occur given that the planet is close to fill its Roche lobe, would lead to an increase of the orbital period producing $O-C$'s  of opposite sign to those observed. 

The measured $O-C$'s of WASP-12, if interpreted as an orbital decay, imply a period decrease of $29 \pm 2$~ms~yr$^{-1}$, meaning a decay timescale for the orbit of only $3.25 \pm 0.23$~Myr and a stellar modified tidal quality factor \citep{Zahn08,MardlingLin02,Ogilvie14} $Q^{\prime}_{\rm s} = 1.8 \times 10^{5}$. Such a fast orbital decay has not been observed in any other similar system. For example, WASP-18  with a planet of $\sim 10$ Jupiter masses on a $\sim 0.94$-day orbit would show a much larger and easily detectable orbital decay for a similar value of $Q^{\prime}_{\rm s}$ {\citep[cf.][]{Maciejewskietal20}}.  Furthermore, the very fast orbital decay of WASP-12 seems to be at variance with the relatively large number of observed HJ systems, although the possibility that we are observing WASP-12 during the very final phase of its life may not be completely excluded \citep[][]{Yeeetal20}. 

Additional information comes from the estimates of $Q_{\rm s}^{\prime}$ for the F-type star in CoRoT-11, having a spectral type similar to WASP-12, that suggest significantly larger values with $ 4 \times 10^{6} \la Q^{\prime}_{\rm s} \la 2 \times 10^{7}$ \citep{Lanzaetal11}.  
A statistical analysis of the HJ population  constrains the most probable  tidal quality factors of their host stars giving $Q^{\prime}_{\rm s}$ between $10^{7}$ and $10^{8}$ \citep[cf.][]{Jacksonetal09,Bonomoetal17,CollierCameronJardine18}, in agreement  with the  dynamical tide theory of \citet{OgilvieLin07} and \citet{Ogilvie14}. Therefore, there is considerable tension between the value of $Q^{\prime}_{\rm s}$ as derived from the alleged orbital decay of WASP-12 and the results coming from the analysis of the HJ population. Unless the system has been caught in a very specific and short phase of its evolution, when resonant g-mode oscillations are excited in the interior of its F-type star, an explanation of the $O-C$'s in terms of a tidal orbital decay encounters significant difficulties \citep[cf.][]{BaileyGoodman19}. 

In view of this conclusion, it is worth proposing and investigating alternative models for the variation of the orbital period in HJ systems, not related to their tidal orbital decay. In the next section, we introduce one of such models that leads to a long-term modulation of the orbital period as a consequence of the gravitational coupling between the orbital motion and the spin of the hot Jupiter produced by a permanent quadrupole deformation of the core of the planet. 

\section{Model}
\subsection{Overview}
\label{overview}
\begin{figure}
\centerline{
\includegraphics[height=6cm,width=9.5cm,angle=0,trim=4.5cm 4cm 4cm 4cm,clip]{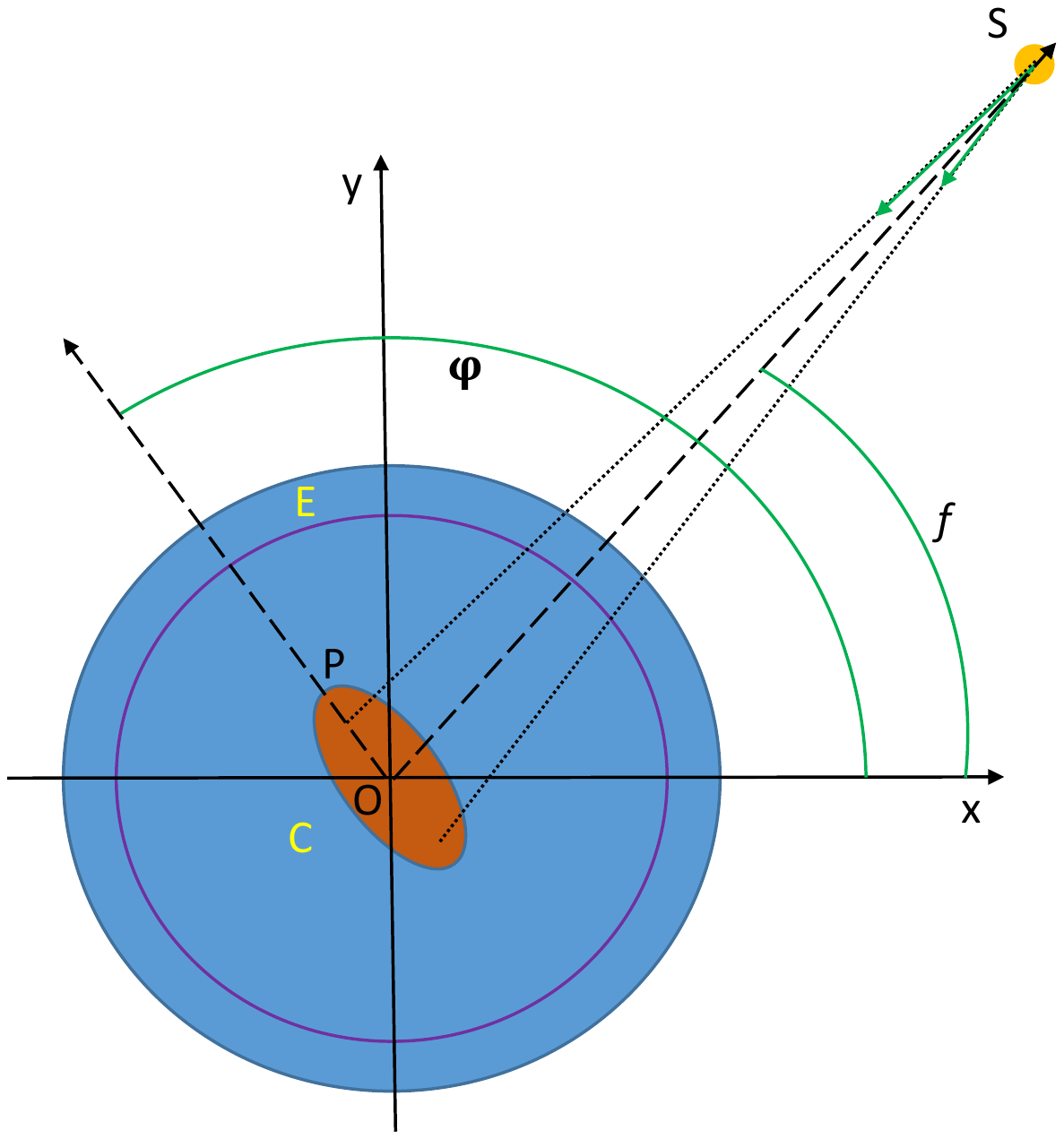}} 
\caption{Sketch of a hot Jupiter planet consisting of a central shell $C$ and an outer shell $E$, separated by the purple boundary, with its host star $S$, treated as a point mass for the purpose of our model. The system is viewed from the North pole of the planet with the observer along its spin axis that coincides with the $\hat{z}$ axis. Inside the central shell $C$, there is a solid and rigid core (in brown) with a permanent non-axisymmetric quadrupole moment resulting from an ellipsoidal deformation, greatly exaggerated in the figure for the purpose of clarity. The barycentre of the planet is indicated with $O$ and is the origin of our Cartesian reference frame with the $\hat{x}$ and $\hat{y}$ axes fixed in an inertial space and located in the equatorial plane  of the planet. The angle between the semimajor axis of the rigid core $OP$ and the $\hat{x}$-axis is indicated with $\varphi$, while the true anomaly of the star is indicated by  $f$. The two green arrows show the gravitational forces acting on the star because of the non-axisymmetric deformation of the planet core. Their resultant has a component tangent to the orbit that leads to an exchange of angular momentum between the planetary core and the orbital motion. }
\label{model_overview1}
\end{figure}

We consider a system consisting of a hot Jupiter and its host star $S$ (see Figure~\ref{model_overview1}). The interior of the hot Jupiter is subdivided into a central shell $C$ and an envelope shell $E$ surrounding it. Inside the central shell, there is a solid and rigid core which has a non-axisymmetric ellipsoidal shape. 

This core may be formed  by the collapse of heavy elements and rocky materials to the centre of the planet during the protoplanetary phase. We assume that the protoplanet came close to its host star during the early stages of its evolution by, for example, type II migration in a protoplanetary disc and that the core solidified soon  when the planet was close to the star \citep[e.g.,][]{DawsonJohnson18}. Therefore, the core was deformed by the stellar tides and acquired a non-axisymmetric ellipsoidal shape that was kept after its solidification  giving it a permanent quadrupole moment. Note that this is the main assumption of our model because the formation and the physical state of the cores of giant planets are presently not known \citep[cf.][]{FortneyNettelmann10}. 

In the case of Jupiter, the measurements of the Juno probe did not found any significant non-axisymmetric quadrupole moment \citep{Iessetal18}, but this result does not invalidate our assumption because Jupiter's core is probably not solid and was formed far from the Sun where any  tidal deformation was negligible. 

A non-axisymmetric quadrupole moment in the core of the planet leads to a non-radial gravitational force acting on the host star that produces a torque transferring angular momentum from the core rotation to the orbital motion and viceversa. A quantitative treatment of the effect is given in Section~\ref{lagrangian_eq}, where we shall derive the equation of the orbital motion in the case of a circular orbit as (cf. equation~\ref{ddotf_eq}):
\begin{equation}
m r^{2} \ddot{f} = \frac{9}{8\pi} \frac{Gm_{\rm s}}{r^{3}} T \sin 2\alpha, 
\label{equa1}
\end{equation}
where  $G$ is the gravitation constant, $f$ the true anomaly, $m_{\rm s}$ the mass of the host star, $r$ the orbital radius, $m \simeq m_{\rm p}$ the reduced mass of the system with $m_{\rm p}$ being the mass of the planet, $T$ the gravitational quadrupole moment of the core as defined in Section~\ref{core_potential}, $\alpha \equiv f -\varphi$ with $\varphi$ being the rotational coordinate of the planet (cf. Figure~\ref{model_overview1}), and a dot over a variable indicates its time derivative. 

According to equation~(\ref{equa1}), an oscillation  of the angle $\alpha$ leads to an oscillation of the true anomaly that in turn produces an oscillation of the $O-C$. This happens because a circular orbit of constant period $P$ corresponds to a constant $\dot{f}=2\pi/P$, that is, $\ddot{f}=0$. The difference in the time of mid-transit in the case of an oscillating true anomaly is given by 
\begin{equation}
O-C = \frac{\Delta f}{2\pi} P,
\label{oc_eq}
\end{equation}
where $\Delta f$ is the difference in the true anomaly at mid transit between the orbit with $\ddot{f} \not=0$ and a constant-period reference orbit with $\ddot{f} =0$, while $P$ is the orbital period of the reference orbit. In conclusion, an oscillation of $\alpha$ produces a modulation of the $O-C$ vs. the time as a consequence of the oscillation of $f$. 

The rotation of the planet is almost perfectly synchronized with its orbital motion owing to the strong tides inside the planet raised by its host star (cf. Sect.~\ref{tides}). Furthermore, the relative amplitude of the modulation of the orbital period $\Delta P/P$ required to explain the $O-C$ diagram of WASP-12 with our model is of the order of $10^{-6}$ over a decade (cf. Sect.~\ref{results_wasp12}), thus $\ddot{f}$ is very small and $\dot{f} \simeq \dot{\varphi}$, that is, a synchronized planet rotation is a very good approximation. This implies that the angle $\alpha$ changes very slowly giving a small quadrupole moment $T$ the time to transfer a sufficient amount of angular momentum between the planet spin and the orbital motion (cf. Figure~\ref{model_overview1}). On the other hand, were the planet rotation away from synchronization with the orbital period, the angle $\alpha$ would vary rapidly and the fast oscillations of ${f}$ would average to zero over a short time interval giving no measurable effect on the $O-C$. 

The total angular momentum of the system is conserved because all the gravitational forces are internal. Considering the host star $S$ as a point mass, that is neglecting any variation of its spin angular momentum, the conservation of the angular momentum can be written as
\begin{equation}
m r^{2} \dot{f} + I \dot{\varphi} = J, 
\end{equation}
where $I$ is the moment of inertia of the rigidly rotating planet and $J$ the total angular momentum of the system that is a constant of the motion. The moment of inertia of the orbit is  $m r^{2}$ and is $\sim 10^{3}$ times larger than the moment of inertia of the planet in very hot Jupiter systems. An orbital period modulation of relative amplitude $\Delta P /P = 10^{-6}$ corresponds to  a relative modulation of the orbital angular velocity of $\Delta \dot{f} / \dot{f} = -\Delta P /P$. This implies a relative variation of the spin angular velocity of the planet by
\begin{equation}
\frac{\Delta \dot{\varphi}}{\dot{\varphi}} = - \frac{mr^{2}}{3I} \frac{\Delta \dot{f}}{\dot{f}} =  \frac{mr^{2}}{3I} \frac{\Delta P}{P},
\end{equation}
where the factor 3 in the denominator in the r.h.s. comes from the variation of the orbital separation during the orbital period modulation computed by means of the Kepler III law (see Sect.~\ref{lagrangian_eq}). This leads to a relative angular velocity modulation of the planet of the order of $\Delta \dot{\varphi}/ \dot{\varphi} \sim 10^{-3}$, in agreement with the slow  variation of the angle $\alpha$ required to transfer a sufficient amount of angular momentum between the orbit and the planet spin in our model. Such a deviation of the planet rotation from synchronism is so small that the timescale for tides to restore a perfect synchronism is of the order of several Myr, much longer than the oscillations of the orbital period, so tides can be neglected in our model (cf. Sect.~\ref{tides} for a justification of this result). 

The above description of our model assumes that the planet rotates rigidly. However, this may not always be the case.  As we shall see (cf. Sect.~\ref{torsional_osc}), {convection in the outer envelope $E$ can occur in a time-dependent regime for some parameters of the planet leading to oscillating Reynolds stresses that impose a variable torque at the outer boundary of the internal $C$ shell (see Figure~\ref{model_overview1}). The $C$ shell rotates rigidly owing to an internal magnetic field that redistributes its angular momentum over a shorter timescale than that of the oscillations of the  Reynolds stresses. Such a rigid rotation is not constant in time, but is modulated at the level of $0.1$ percent by the time-variable Reynolds stresses. The ellipsoidal solid core is at the centre of the $C$ shell and is assumed to be rigidly coupled to the rotation of the shell itself by the internal magnetic field.

In conclusion, regarding the internal rotation of the planet as a whole, we can have two different regimes. In the first regime,  the Reynolds stresses in the outer envelope $E$ show small fluctuations in time and  the whole planet rotates almost as a rigid body, therefore the angular momentum exchanged with the orbital motion is redistributed almost uniformly over the whole planetary interior by the magnetic (in the $C$ shell) and the Reynolds stresses (in the $E$ shell). In the second regime, the oscillations of the Reynolds stresses in the $E$ shell have a remarkable amplitude and the amount of angular momentum exchanged between the $C$ and $E$ shells is comparable with or larger than the angular momentum exchanged with the orbit. In this case, the internal rotation of the planet cannot be treated as rigid and the exchange of angular momentum between the $C$ and $E$ shells must be included explicitly into the model for the orbital period variation. We discuss these two regimes in detail in Sects.~\ref{torsional_osc} and~\ref{lagrangian_eq}. Here we anticipate that the amount of angular momentum that can be exchanged between the planet and the orbit can be larger in the latter regime leading to orbital period modulations of larger amplitude than in the former regime. }

After giving this qualitative overview of our model, we now consider in details its main ingredients and provide justifications for the adopted hypotheses. 

\subsection{The role of tides}
\label{tides}
Tides raised by the host star inside the close-by planet lead to the synchronization of planet rotation with the orbital motion and the alignment of its spin with the orbital angular momentum. They also damp any initial eccentricity of the orbit on a timescale  much shorter than the main-sequence lifetime of the system, so we can assume that the orbit is circular \citep[cf.][]{Ogilvie14}.  

An estimate of the timescale $\tau_{\rm ps}$ over which tides  synchronize the planet spin with the orbital motion can be derived by eq.~(9) of \citet{Guetal03}:
\begin{equation}
\tau_{\rm ps}^{-1} \equiv \frac{\dot{\Omega}_{\rm p}}{n-\Omega_{\rm p}} = \frac{9 n}{2h_{\rm p} Q^{\prime}_{\rm p}} \left( \frac{m_{\rm s}}{m_{\rm p}} \right) \left( \frac{R_{\rm p}}{r} \right)^{3},
\label{tidal_timescale}
\end{equation}
where $\Omega_{\rm p} = \dot{\varphi}$ is the angular velocity of the planet, $n=2\pi/P$ the orbital mean motion, $Q_{\rm p}^{\prime}$ the modified tidal quality factor of the planet, $h_{\rm p} =I/(m_{\rm p} R_{\rm p}^{2})$ the normalized moment of inertia of the planet, $I$ the planet moment of inertia, $m_{\rm p}$ the mass of the planet, $R_{\rm p}$ its radius, $m_{\rm s}$ the mass of the star,  and $r$ the radius of the circular orbit. 

The tidal quality factor $Q_{\rm p}^{\prime}$ is a function of the tidal frequency $\hat{\omega}$. In the case of a circular orbit, the semidiurnal tide is dominant, so we can assume $\hat{\omega} = 2(n-\Omega_{\rm p})$. The dependence of $Q^{\prime}_{\rm p}$ on $\hat{\omega}$ can be very complex and characterized by rapid oscillations and resonances when the tidal potential excites oscillations inside the planet \citep{OgilvieLin04}. Presently, tidal theory is not advanced enough to predict the value of $Q_{\rm p}^{\prime}$ from first principles, so we can estimate it by considering observations of the system of the Galileian moons of Jupiter \citep{Laineyetal09}  giving a value of $\sim 10^{5}$ for Jupiter when the tidal frequency $\hat{\omega}$ is comparable with the rotation frequency of the planet. 

In the case of HJs, the planet is very close to synchronization, thus the tidal frequency $\hat{\omega} \ll \Omega_{\rm p}$. We do not have observations that  constrain $Q_{\rm p}^{\prime}$ in this regime. Note that the estimates obtained from the eccentricity of HJ orbits by, e.g., \citet{Bonomoetal17} refer to the $Q^{\prime}_{\rm p}$ associated with the annual tides, not the semidiurnal tides considered here that have a much smaller frequency when the planet is close to synchronization. Therefore, we may only extrapolate from the value obtained for Jupiter making some theoretical assumptions, for example, assuming that the time lag between the tidal potential and the tidal bulge is approximately constant which leads to a dependence of the form $Q^{\prime}_{\rm p} \propto \hat{\omega}^{-1}$ \citep[cf.][for a discussion of this assumption]{Leconteetal10,Eggletonetal98}.  
The level of asynchronism required by our model is of the order $\hat{\omega} \sim 10^{-3} \Omega_{\rm p}$ (cf. Section~\ref{overview});  considering that Jupiter is rotating $2-3$ times faster than our HJs and that  $Q_{\rm p}^{\prime} \propto \Omega_{\rm p}^{2}$ as suggested by \citet{OgilvieLin07}, these imply $Q_{\rm p}^{\prime} \sim 10^{8}-10^{9}$. {The proportionality of $Q^{\prime}_{\rm p}$ to $\Omega_{\rm p}^{2}$ comes from the proportionality of the tidal dissipation to the square of the amplitude of the inertial waves that are responsible for most of the dissipation in the considered tidal regime. They have the Coriolis force as their restoring force, thus their amplitude is proportional to $\Omega_{\rm p}$ \citep[cf.][Section~4.5]{Ogilvie13}.} With the above range for $Q^{\prime}_{\rm p}$ and considering as typical values for our very close planets $P \sim 1$~day, $m_{\rm s}/m_{\rm p} \sim 10^{3}$, and $R_{\rm p}/r \sim 2 \times 10^{-2}$, Equation~(\ref{tidal_timescale}) gives $\tau_{\rm ps} \sim (10^{8} -10^{9}) P$, where we have assumed $h_{\rm p} = 0.26$ that is a good estimate for Jupiter \citep{Nettelmannetal12} and probably also for HJs \citep{Guetal03}. 

We conclude that the tidal timescale to reach synchronization in our HJs is of the order of $1-10$~Myr, that is, much longer than the decadal timescales considered for the orbital period modulation in our model. Therefore, we can separate the effects of tides from the shorter-term dynamics of our system, which greatly simplifies our treatment. 

Tides inside the star raised by the planet tend to synchronize stellar rotation and indeed there are a few HJ systems where this synchronization has been apparently reached, notably in $\tau$~Bootis \citep{Borsaetal15}. In this case, a permanent quadrupole moment inside the star can contribute to the modulation of the orbital period  by the same mechanism introduced in Sect.~\ref{overview}. This possibility has been explored by \citet{Lanza20} to whom we refer the interested reader, but it will not be considered here because most of the stars hosting HJs rotate much slower than the orbit so the effect of their permanent quadrupole moment, if any, is averaged to zero and does not contribute to the orbital period modulation of the system.    

\subsection{The permanent quadrupole moment of the planetary core}
\label{core_potential}

Let us consider a reference frame with the origin in the barycentre $O$ of the planet and the polar axis $z$ coincident with its spin axis. Let $r$ be the distance from the origin, $\theta$  the colatitude measured from the North pole, and  $\phi$  the azimuthal coordinate. The core of the planet is not axially symmetric, but has a permanent ellipsoidal deformation. We assume that the spin axis coincides with one of the principal axes of inertia of the core, while the two other principal axes in its equatorial plane are along the Cartesian $x$ and $y$ axes. 
This reference frame is not an inertial frame, but rotates with the planet, thus we indicate the azimuthal coordinate with $\phi$ to distinguish it from the azimuthal coordinate $\varphi$ in the inertial reference frame considered in Section~\ref{overview}. 

The outer gravitational potential of the planet $U(r, \theta, \phi)$ satisfies the Laplace equation $\nabla^{2} U = 0$ and can be expanded into a series of orthonormal complex spherical harmonic functions:  
\begin{equation}
Y_{lm}(\theta, \phi) = (-1)^{\frac{m+|m|}{2}} j^{l} \left[\frac{2l+1}{4\pi} \frac{(l-|m|)!}{(l+|m|)!}\right]^{\frac{1}{2}} P_{l}^{|m|} (\cos \theta) \exp(jm\phi),
\end{equation}
 where $P_{l}^{|m|}(x)$ is the associated Legendre function of degree $l$ and azimuthal order $m$ with $ -l \leq m \leq l$, and $j=\sqrt{-1}$. The coefficients of the series expansion depend on the relative distance from the origin $r/R_{\rm p}$, where $R_{\rm p}$ is the mean radius of the planet. The terms corresponding to $l=1$ vanish when the origin of the reference frame is chosen in   the barycentre of the body, therefore
\begin{equation}
U (r, \theta, \phi) = -\frac{Gm_{\rm p}}{r} \Re \left\{1 + \sum_{l=2}^{\infty} \left( \frac{r}{R_{\rm p}}\right)^{-l} \sum_{m=-l}^{l} u_{lm} Y_{lm}(\theta, \phi) \right\},
\end{equation} 
where $\Re\{z\}$ is the real part of the complex quantity $z$ and the complex coefficients $u_{lm}$ depend on the density distribution  inside the planet according to:
\begin{equation}
u_{lm} = \frac{1}{(2l+1) m_{\rm p} R_{\rm p}^{l}}\int_{V} (r^{\prime})^{l} Y^{*}_{lm}( \theta^{\prime}, \phi^{\prime}) \rho(r^{\prime}, \theta^{\prime}, \phi^{\prime})\, dV^{\prime}
\label{eq_ucoeff}
\end{equation}
where the asterisk denotes  complex coniugation, $\rho(r, \theta, \phi)$ is the internal density, and the integration is extended over the volume $V$ of the planet \citep[cf.][]{Iessetal18}. Note that $(-1)^{l-m} \,Y_{l,-m} = Y^{*}_{l, m}$.

We assume that the star orbits in the equatorial plane of the planet, therefore, we are interested in the outer gravitational potential in the equatorial plane, that is, we fix $\theta = \pi/2$. We develop the potential up to the terms with $l=2$ because the terms of higher orders decay rapidly as $(r/R_{\rm p})^{-l}$ becoming negligible at the distance of the star. The spherical harmonic $Y_{20} \propto (1-3\cos^{2} \theta)$ and reduces to a term independent of the azimuthal coordinate $\phi$ in the equatorial plane, while $Y_{2, \pm 1} \propto \sin \theta \cos \theta$ vanish in the equatorial plane; thus, only $Y_{2, \pm 2}$ give contributions depending on $\phi$ in the equatorial plane. 

In conclusion, the expression of the outer gravitational potential of the planet in the orbital plane is:
\begin{equation}
U(r, \frac{\pi}{2}, \phi) = -\frac{Gm_{\rm p}}{r} - \frac{Gm_{\rm p}}{r} \left( \frac{R_{\rm p}}{r} \right)^{2} \left[ u_{20} Y_{20} + 2 \Re\{ u_{22} Y_{22} \} \right]
\end{equation}
The solid spherical harmonics $r^{2} Y_{2,m}$ can be expressed in terms of the Cartesian coordinates, thus equation~(\ref{eq_ucoeff}) can be used to express the coefficients $u_{2m}$ in terms of the tensor of inertia of the planet defined by:
\begin{equation}
I_{ij} \equiv \int_{V} \rho ({\bf x})\, x_{i} x_{j} \, dV,
\end{equation}
where ${\bf x} = (x_{1}, x_{2}, x_{3})$ is the position vector and $x_{i}$ with $i=1,2,3$ are the Cartesian coordinates. In the adopted reference frame with the coordinate axes directed along the principal axes of inertia of the core,  only the principal moments of inertia $I_{xx}, I_{yy}, I_{zz}$ are different from zero. The expression for the outer gravitational potential of the planet  becomes:
\begin{eqnarray}
\lefteqn{U(r, \frac{\pi}{2}, \phi)  =  -\frac{Gm_{\rm p}}{r} } & & \nonumber  \\ 
& & - \frac{G}{16\pi r^{3}} \left[(I_{xx} +I_{yy}-2I_{zz}) +3(I_{xx} -I_{yy}) \cos 2 \phi \right]  
\label{eq_outer_pot}
\end{eqnarray}
Introducing the quadrupole moment tensor 
\begin{equation}
Q_{ik} \equiv I_{ik} - \frac{1}{3} \delta_{ik} {\rm Tr} I,
\label{qdef}
\end{equation}
where $\delta_{ik}$ is the Kronecker delta and ${\rm Tr} I = I_{xx} + I_{yy} + I_{zz}$ is the trace of the inertia tensor, we can recast the above expression as:
\begin{equation}
U(r, \frac{\pi}{2}, \phi) = -\frac{GM_{\rm p}}{r}  - \frac{3G}{2r^{3}} \left [ \frac{1}{8\pi} \left( Q_{0} + T \cos 2 \phi \right) \right],
\label{our_quadrupole}
\end{equation}
where we made use of the fact that the tensor $Q$ is traceless, i.e., $Q_{xx} + Q_{yy} + Q_{zz}=0$ as follows from its definition in equation~(\ref{qdef}), and defined $Q_{0} \equiv Q_{xx} + Q_{yy}$ and $T \equiv Q_{xx} - Q_{yy}$.  

We assume that the core of the planet is rigid and has a permanent quadrupole moment acquired when it solidified close to the star (cf. Section~\ref{overview}). In other words, the quadrupole moment was induced by the tidal deformation produced by the star and became permanent when the core became solid. {We can express the quadrupole moment $T$ in terms of the perturbing tidal potential produced by the star on the core when it solidified. This is possible because the external quadrupole potential produced by the core is linearly related to the quadrupole component of the perturbing potential via an appropriate Love number \citep{Ogilvie14}. We use the model by \citet{Remusetal12} that considers an idealized planet consisting of a solid core of uniform density $\rho_{\rm c}$, mass $m_{\rm c}$ and mean radius $R_{\rm c}$, upon which there is a fluid layer of uniform density $\rho_{\rm o}$ and outer mean radius $R_{\rm p}$. By comparing our expression for the quadrupole potential (equation~\ref{our_quadrupole}) with their equation~(49) at the surface $r=R_{\rm c}$ of the core, we find:
\begin{equation}
T = 4 \pi F k_{2} \left(\frac{m_{\rm s}}{m_{\rm c}} \right) \left(\frac{R_{\rm c}}{a} \right)^{3} m_{\rm c} R_{\rm c}^{2},
\label{quadrupole_deformation}
\end{equation}
where $k_{2}$ is the second-order Love number, $m_{\rm s}$ the mass of the host star, $a$ the star-planet separation when the core solidified, and $F$  a function of the density ratio $\rho_{\rm c}/\rho_{\rm o}$, the radius ratio $R_{\rm c}/R_{\rm p}$,  and the  effective rigidity of the core as given by equations~(27), (54), and~(56) of \citet{Remusetal12}. We define $k_{2} \equiv (3/2)/(1+ \bar{\mu})$, where $\bar{\mu}$ is the effective rigidity of the core \citep[cf. equation~27 of][]{Remusetal12}. The factor $F$ takes into account the modification of the core deformation due to the effects of the external fluid layer the weight of which acts on the core and which is also deformed by the tidal potential of the star. In the case of a naked core, $F=1$, while it increases in the case of a deep envelope ($R_{\rm c}/R_{\rm p} \la 0.6$) reaching a limiting value of about 2.3 \citep[cf. Figure~5 in][]{Remusetal12}. Note that an increase of $F$ beyond the unity implies a stronger deformation of the core for a given mass $m_{\rm c}$ and radius $R_{\rm c}$ as a consequence of the weight of the tidally distorted fluid envelope. 

For a giant planet the value of the Love number $k_{2} \sim 0.36$ from a model of its interior stratification; a slightly smaller $k_{2}$ is found in the case of a solid planet, for example, $k_{2} = 0.295$ for the Earth \citep{Lainey16}, while for a homogeneous fluid body $k_{2} = 3/2$ \citep{Remusetal12,Ogilvie14}.  }

\subsection{Angular momentum transport inside the planetary interior}
\label{torsional_osc}
The permanent quadrupole moment of the rigid core produces an exchange of angular momentum with the orbital motion as we shall see in detail in Section~\ref{lagrangian_eq}. The exchanged angular momentum is then redistributed from the core to the whole planetary interior by different physical mechanisms. In this Section, we investigate such mechanisms because they play a crucial role in our model. 

Molecular viscosity inside giant planets can be neglected for our purposes. In the case of the Jupiter model computed by \citet{Frenchetal12}, the kinematic viscosity $\nu$ in the planet's interior is in the range $(0.25-0.5) \times 10^{-6}$~m$^{2}$~s$^{-1}$ giving a characteristic timescale for the angular momentum transport across the planet's radius of $R_{\rm J}^2/\nu \sim (3-6) \times 10^{14}$ yr, where $R_{\rm J}$ is the radius of Jupiter. 

The turbulent viscosity produced by internal convective motions can be estimated as $\nu_{\rm turb} = (1/3) \alpha_{\rm ml} H_{\rm p} u_{\rm c}$, where $\alpha_{\rm ml} \sim 1.5$ is the ratio of the mixing length to the local pressure scale height $H_{\rm p}$ and $u_{\rm c}$ the convective velocity. The latter can be estimated in the mixing-length theory from the flux transported by convection $F_{\rm c}$ as $u_{\rm c} \simeq (0.1 \alpha_{\rm ml} F_{\rm c}/\rho)^{1/3}$, where $\rho$ is the density \citep{Lanza05,Kippenhahnetal12}. In the case of Jupiter, \citet{Jones14} estimates convective velocities between $10^{-3}$ and $10^{-2}$~m~s$^{-1}$ giving a diffusion timescale of the angular momentum across the whole planet of $\sim 4\times 10^{4}$ yr. However, the convective turbulent velocities inside HJs may be higher if some small fraction of the stellar insolation is conveyed into the deep interior as considered in some models proposed to explain the inflated radii observed in a sizable fraction of giant close-by planets \citep[e.g.,][]{GuillotShowman02,Laughlin18,Sainsburyetal19}. For example, assuming that 1~percent of the insolation received by WASP-12 is conveyed to the planet deep interior, it is possible to account for its inflated radius \citep[e.g.][]{GuillotHavel11}. If such an excess flux is transported by convection, turbulent velocities up to $\sim 10$~m~s$^{-1}$ are required giving a larger turbulent viscosity that implies characteristic turbulent diffusion timescales of the order of $10^{2}$ yr. We conclude that, even in this extreme case, the turbulent transport of angular momentum inside a giant planet is slow in comparison with orbital period variations occurring on timescales of decades. Therefore, we look for other processes to produce a faster internal exchange of angular momentum in HJs, notably those associated with an internal magnetic field.  

Giant planets in the solar system host internal hydromagnetic dynamos that produce magnetic fields with intensities up to tens of Gauss at the surface \citep[e.g.,][]{RuedigerHollerbach04,Jones14}. The low viscosity and the nearly polytropic stratification \citep{Frenchetal12} together with the fast rotation make the internal angular velocity constant along cylinders co-axial with the rotation axis according to the so-called Taylor-Proudman regime \citep{RuedigerHollerbach04}. Each co-axial cylinder can rotate with a different angular velocity with the magnetic field providing Maxwell stresses that couple different cylinders in the region of high electric conductivity in the planetary interior. This hydromagnetic system can develop torsional oscillations as discussed by, e.g., \citet{Horietal19}. These oscillations can be excited by the azimuthal component of the Lorentz force that is generally different from zero in hydromagnetic dynamos \citep{RuedigerHollerbach04} and have been considered as a mechanism to couple the inner solid core of the Earth with its external fluid core and the mantle \citep{RobertsAurnou12} to explain the cyclic variation of the length of the day with a relative amplitude of $\approx 2 \times 10^{-8}$. 

{The amplitude of the torsional oscillations in planetary dynamo models is very small in comparison with the angular velocity of rotation. In other words, they represent a small perturbation with respect to a state of rigid rotation  that is enforced in the conductive interior by the angular momentum transported by the waves themselves on timescales of the order of $\la 10$~yr, shorter than the modulation cycles of the orbital period in our model \citep{WichtChristensen10,Horietal19}. In conclusion, the hydromagnetic dynamo is capable of enforcing an almost rigid rotation up to the level where the electric conductivity decreases because of the transition of the hydrogen and helium from an ionized to a molecular state. In the case of Jupiter, such a decrease of the conductivity happens between 0.85 and 0.9 of its radius \citep{GastineWicht12,Wichtetal18} and a similar relative radius can be assumed for hot Jupiters \citep[e.g.,][]{BatyginStevenson10}.

The outer shell of the planet with a low electric conductivity can be approximately modelled in a purely hydrodynamic way, neglecting the effects of the magnetic field because it decouples from the flow. In this shell, energy is transported by convection up to the radiative atmosphere of the planet. The dynamical coupling between this outer convective shell and the rigidly rotating planetary interior has been modelled by \citet{HeimpelAurnou12}, in an attempt to connect possible variations in the internal rotation of Saturn to the temporal variability in the zonal winds of its atmosphere \citep[cf., e.g.,][Sect.~2.4]{Wichtetal18}, and we shall refer to their model for our purposes.

In rapidly rotating giant planets,  zonal flows are maintained by the equatorward transport of angular momentum mainly by turbulent Reynolds stresses with the meridional circulation playing a minor role \citep{Wichtetal18}. Therefore, a time-variable convection, leading to variable Reynolds stresses, can produce changes in the speed of the zonal flows. \citet{HeimpelAurnou12} model convection in a rotating shell considering the Boussinesq approximation, that is, assuming an incompressible fluid, and find a regime characterized by cyclic bursts of convection leading to a remarkable amplification of the zonal flows followed by a slow decay before the next burst. The angular momentum transported by the Reynolds stresses to accelerate the prograde equatorial zonal flow leads to a braking of the rotation of the interior because of the conservation of the total angular momentum of the planetary rotation. In the specific model they consider, the cyclic oscillations of the rotation of the planetary interior produced as a consequence of the convective bursts have an amplitude of about 0.1 percent. 

The role of compressibility  has been explored by \citet{GastineWicht12} who also considered the effects of varying the Rayleigh number of the hydrodynamic simulations, while keeping the shell aspect ratio and the Ekman number fixed. They found that the convection in the outer shell is characterized by different regimes with the transitions from one to the other controlled by the Rayleigh number and the density stratification. Vacillating and intermittent regimes of convection, with regular  oscillations of the kinetic energy of the zonal flows, similar to the simulations of \citet{HeimpelAurnou12}, are found together with chaotic regimes characterized by fluctuations of the kinetic energy of the flow as large as tens of percents in some cases or of less than 1 percent in others, the latter corresponding to nearly stationary rotation. 
By increasing the stratification and the Rayleigh number, the vacillating regime tends to disappear, but the intermittent and the chaotic regimes are found up to the borders of the explored parameter space. 
Although \cite{GastineWicht12} do not model the angular momentum exchange between the convective shell and a rigidly rotating interior, their results support and extend the conclusions of \cite{HeimpelAurnou12}. More precisely, they show that two general convection behaviours are possible in the outer shell, that is, one associated with large temporal variations of the rotation with amplitudes up to $\sim 10-15$ percent, both in a cyclic or in a chaotic way, and another characterized by an almost stationary rotation. 

Following the work by \citet{Ballotetal07}, \citet{Wichtetal18} interpret the cyclic variation of the rotation in the convective shell as the result of a competition between convection and shear. At the beginning of the cycle, the radial shear is low and the convective plumes are almost undisturbed and produce Reynolds stresses leading to a transport of angular momentum towards the equator and the upper boundary of the shell under the action of the Coriolis force. This produces a steady increase of the radial shear across the shell until, when the shear exceeds a critical amplitude, convective plumes are disrupted. At this point, Reynolds stresses become negligible and the shear across the shell is progressively reduced by turbulent eddy viscosity until it becomes so small that convective plumes can resume and start a new cycle. During the chaotic regime, these oscillations become aperiodic, but variations in the rotation of amplitude up to $10-15$ percent are still possible in some domains of the parameter space, depending on the duration of the shear-dominated and convection-dominated phases  in comparison with the turbulent diffusion timescale across the shell. In such a case, the angular momentum exchanged with the planetary interior can lead to variations of its rotation with amplitudes of the order of 0.1 percent because its moment of inertia is remarkably larger than that of the convective outer shell. On the other hand, for other values of the characteristic parameters,  the amplitudes of the rotation fluctuations become small, the angular momentum exchanged with the interior negligible, and the whole planet rotates almost rigidly. 

The numerical simulations currently available do not completely sample the full accessible parameter space \citep{Wichtetal18}. In any case, the hydrodynamic regimes of real planetary convective envelopes are many orders of magnitudes away in terms of characteristic parameters from the regimes accessible to numerical simulations, thus we cannot directly apply these results to them. 

A scaling of the results by \citet{HeimpelAurnou12} on the basis of the Ekman number shows that the periods of the convective cycles in real giant planets should be of the order of decades, while the amplitude of the rotation variations should not strongly depend on the Rayleigh number $Ra$ because the Reynolds stresses do not depend on $Ra$ when we extrapolate according to the asymptotic regime displayed by the simulations of \citet{GastineWicht12}. Note that the timescale of the variation in the internal rotation is much shorter than the $1-10$~Myr timescale typical of tidal angular momentum exchanges (cf. Section~\ref{tides}). Therefore, we can neglect tidal effects on the time-dependent rotation considered in our model. 

In conclusion, we can assume that the internal rotation of our hot Jupiters can occur in one of two different regimes: a) planets with an almost time-independent internal rotation, corresponding to a regime characterized by  little exchange of angular momentum between the interior  and the envelope  because of low-amplitude fluctuations in the envelope convective motions; b) planets with an internal time-dependent rotation produced by cyclic exchanges of angular momentum between the interior  and the envelope  because of a variable shear at the base of the envelope where convection is intermittent or vacillating. Aperiodic fluctuations of the rotation of comparable amplitude may also occur in regimes of chaotic convection in the envelope. 

In the planets characterized by the latter regime, there is a transition layer at the base of the convective envelope that produces a time-variable torque $\Gamma (t)$ on the interior whose rotation is maintained rigid by the dynamo field (see above). Such a torque can be regarded as periodic in the case of intermittent or vacillating convection.   We refer to \citet{HeimpelAurnou12} for a discussion of the properties of the transition layer, in particular of its thickness and location, because what is relevant here is the constancy (case a) or the cyclic oscillation (case b) of the rotation of the interior of the planet. The amplitude of the cyclic oscillations of the interior rotation can be assumed to be of the order of 0.1 percent.

It is interesting to note that a variability of the differential rotation has been observed in some rapidly rotating late-type stars as well. For example, the late G dwarf AB Doradus with a rotation period of $\sim 12$~hr, has shown variations of its equatorial angular velocity with a relative amplitude of  $\sim 0.004$ over a time span of $\sim 8$ years \citep{CollierCameronDonati02,Lanza06}. \citet{Ballotetal07} suggested that such  oscillations could be the results of time-dependent Reynolds stresses produced by a vacillating convection. However, the matter is significantly ionized throughout the stellar interior, thus we cannot apply a purely hydrodynamic model to simulate the outer convection zone of AB Dor as in the case of an hot Jupiter \citep{Wichtetal18}. }

In our simplified model, in the case of the regime b), we consider that the angular momentum is exchanged between the two internal shells $C$ and $E$ (cf. Figure~\ref{model_overview1}) that we assume to be cylindrical shells co-axial with the planet spin axis separated by a cylindric boundary of radius $s_{0}$ because the internal rotation is in the Taylor-Proudman regime. The moment of inertia of the inner shell $C$ is indicated with $I_{\rm c}$, while that of the outer shell $E$ is $I_{\rm e}$. The inner shell contains the solid core of the planet with its permanent quadrupole moment and is strongly coupled to it as to rotate with the same angular velocity. 

\begin{figure}
\centerline{
\includegraphics[height=10.5cm,width=7.5cm,angle=90]{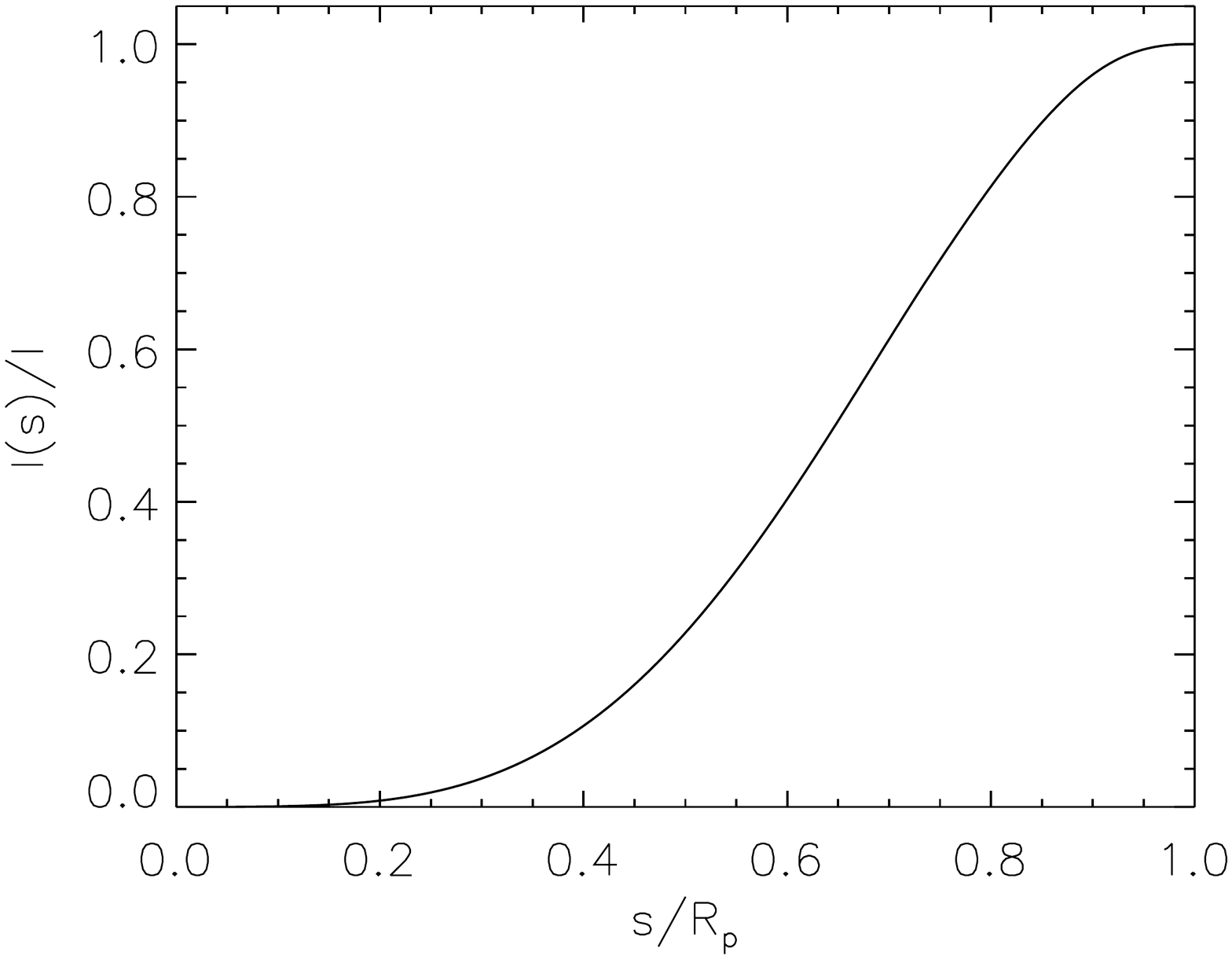}} 
\caption{Ratio of the moment of inertia $I(s)$ of the part of Jupiter inside the cylindrical radius $s$ to the total planet moment of inertia $I$ vs. the relative radius $s/R_{\rm p}$. For reference, $I = 2.52 \times  10^{42}$~kg~m$^{2}$ and $R_{\rm p} = R_{\rm J} = 7.14 \times 10^{7}$~m.  }
\label{inertia_model}
\end{figure}

Assuming the internal structure model of Jupiter by \citet{Nettelmannetal12} can be scaled to HJs, we plot in Figure~\ref{inertia_model} the ratio $I(s)/I$ vs. the relative radius $s/R_{\rm p}$, where $I(s)$ is the moment of inertia of the part of the planet interior to the cylindric radius $s$ and $I$  its total moment of inertia. We see that the {moment of inertia of the shell $E$ above a radius $s_{0} = 0.9$~$R_{\rm p}$ is $\sim 5$ percent of the moment of inertia of the rigidly rotating interior $C$ of the planet, that is, $I_{\rm e} \sim 0.05 \, I_{\rm c}$. We recall that $s_{0} \sim 0.9$~$R_{\rm p}$ is determined by the transition from metallic to molecular hydrogen and helium in the interior. }

\subsection{Equations of motion}
\label{lagrangian_eq}

\begin{figure}
\centerline{
\includegraphics[height=6.0cm,width=9.5cm,angle=0,trim=4.5cm 3cm 4.5cm 4.8cm,clip]{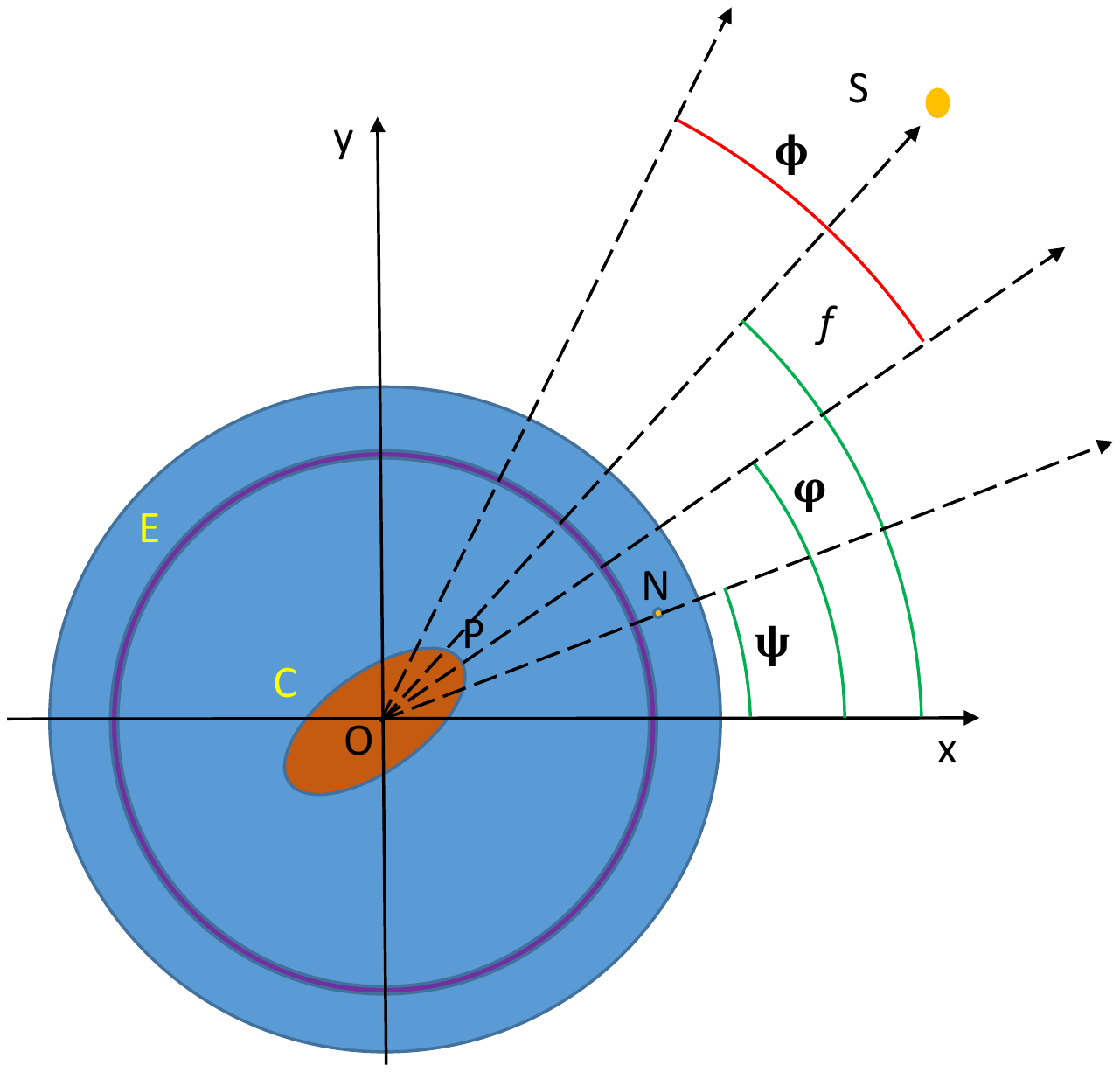}} 
\caption{Specification of the angular coordinates used in the Lagrangian function of our system in a reference frame with the origin at the barycentre of the planet $O$. The Cartesian axes $\hat{x}$ and $\hat{y}$ are fixed in an inertial space and lie in the equatorial plane of the planet; the $\hat{z}$ axis is directed along the planet spin and is pointing towards the observer. The star is treated as a point mass located in $S$. The rotational coordinate of the planetary solid core $\varphi$ is the angle between the $\hat{x}$-axis and its principal axis of inertia directed along its major axis. The angle $\varphi$ is used also as the rotational coordinate of the internal shell $C$. The rotational coordinate $\psi$ of the outer shell $E$  is measured from the $\hat{x}$-axis to a reference point $N$ in the shell itself, defined by averaging over the fluctuating convective motions. The true anomaly of the orbital motion $f$ is  measured from the $\hat{x}$-axis because the orbit is assumed to be circular, so the direction of the periapsis is arbitrary. For completeness, the angle $\phi$, measured from the principal semiaxis $OP$ of the core and introduced in Section~\ref{core_potential}, is also indicated. It is marked with a red arc to make it clear that it is not measured from the $\hat{x}$-axis as the other angles that are marked by green arcs. }
\label{model_frame}
\end{figure}

To write the equations of motion of our star-planet system, we first write the expression of its Lagrangian function 
\begin{equation}
{\cal L} \equiv {\cal T} - {\Psi},
\end{equation}
 where $\cal T$ is the total kinetic energy and $\Psi$ the potential energy of the system expressed as functions of the coordinates and their time derivatives in an inertial reference frame \citep[e.g.,][]{Goldstein50}.  We choose the origin of the reference frame in the barycentre of the star-planet system $Z$ and write the total kinetic energy as the sum of the energy of the orbital motion of the star and the planet around $Z$ and their kinetic energy of rotation around their own barycentres $S$ and $O$, respectively. For simplicity, we regard the star as a point mass, thus neglecting its kinetic energy of rotation, and introduce the reduced mass of the system $m \equiv m_{\rm s} m_{\rm p}/(m_{\rm s} + m_{\rm p})$ to simplify the expression of the kinetic energy of the orbital motion around $Z$, where $m_{\rm s}$ and $m_{\rm p}$ are the mass of the star and the planet, respectively. To express the kinetic energy of rotation of the planet, we adopt the reference frame defined in Figure~\ref{model_frame}, the origin of which is at the barycentre $O$ of the planet, while the axes $x$ and $y$ are fixed in an inertial space and lie in the equatorial plane of the planet that coincides with the orbital plane of the system.  
 
The distance $OS$ between the planet and the star is indicated by the radial coordinate $r$, the true anomaly of the relative orbit is $f$, the angle of the principal major axis of inertia of the planetary core with the $x$-axis is $\varphi$ and is used to measure the rotation of the inner shell $C$, while the angle measuring the rotation of the planetary outer shell $E$ is $\psi$. This is measured with respect to a reference point of the cylindrical shell $E$ that is assumed to rotate with a {mean angular velocity} $\dot{\psi}$ (cf. Figure~\ref{model_frame}). Similarly, the cylindrical shell $C$, including the core, is assumed to rotate rigidly with the angular velocity $\dot{\varphi}$. 

The expression of the kinetic energy when the spin and orbital angular momenta are aligned is:
\begin{equation}
{\cal T} = \frac{1}{2} m \left( \dot{r}^2 + r^{2} \dot{f}^2 \right) + \frac{1}{2} I_{\rm c} \dot{\varphi}^{2} + \frac{1}{2} I_{\rm e} \dot{\psi}^{2},
\end{equation}
where  $I_{\rm c}$ and $I_{\rm e}$ are the moments of inertia of the $C$ and $E$ shells, respectively. 

The potential energy of the system consists of the gravitational energy and {the energy associated with the work done by the torque $\Gamma(t)$ that produces an exchange of angular momentum between the $C$ and $E$ shells.} Using the results obtained in Sections~\ref{core_potential} and~\ref{torsional_osc} and considering that the orbit lies in the equatorial plane of the planet, we write its expression as:
\begin{equation}
\Psi = -\frac{Gm_{\rm s} m_{\rm p}}{r} -  \frac{3Gm_{\rm s}}{2 r^{3}} \left[ \frac{1}{8\pi} \left( Q_{0} + T \cos 2\alpha \right) \right] + \Gamma(t) (\psi - \varphi) , 
\end{equation}
where all the quantities have been introduced in the above Sections and we define the angle $\alpha \equiv f - \varphi$ (cf. Figure~\ref{model_frame}).  Note that the torque $\Gamma (t)$ is taken positive when it accelerates the rotation of the inner shell $C$ and is in general a function of the time. 

Applying the Lagrangian formalism, we derive the following equations of motion:
\begin{equation}
\begin{array}{l}
{\displaystyle \ddot{r} - r \dot{f}^{2} + \frac{Gm_{\rm t}}{r^{2}} + \frac{9Gm_{\rm t}}{16 \pi m_{\rm p} r^{4}} \left( Q_{0} +  T \cos  2 \alpha \right) = 0, }\\
~\\
{\displaystyle \frac{d}{dt} \left( m r^{2} \dot{f} \right) +  \frac{3Gm_{\rm s} T}{8\pi r^{3}} \sin 2 \alpha = 0. } \\
~\\
{\displaystyle I_{\rm c}  \ddot{\varphi} - \frac{3Gm_{\rm s} T}{8\pi r^{3}} \sin 2 \alpha - \Gamma(t) = 0, } \\
~\\
{\displaystyle I_{\rm e} \ddot{\psi} + \Gamma (t) = 0, } 
\end{array}
\label{eqs_motion}
\end{equation} 
where $m_{\rm t} \equiv m_{\rm s} + m_{\rm p}$ is the total mass of the system. From the first of equations~(\ref{eqs_motion}), considering that $\ddot{r}=0$ in the case of a circular orbit, we derive a generalized expression for the Kepler III law  as
\begin{equation}
r^{3}\dot{f}^{2} = Gm_{\rm t} \left[1 + \frac{9}{16\pi m_{\rm p} r^{2}} \left(Q_{0} + T \cos 2 \alpha \right) \right] \simeq Gm_{\rm t},
\label{kepler3}
\end{equation}
where the last equality follows from the smallness of the quadrupole moment terms $Q_{0}$ and $T$ in comparison with the moment of inertia of the orbit $m_{\rm p} r^{2}$. 

The conservation of the total angular momentum of the system follows by summing together the last three of equations (\ref{eqs_motion}) and integrating with respect to the time:
\begin{equation}
m r^{2} \dot{f} + I_{\rm e} \dot{\psi} + I_{\rm c} \dot{\varphi} = J,
\label{amc}
\end{equation}
where $J$ is the total angular momentum of the system. By means of the Kepler III law $r^{3} \dot{f}^{2} = Gm_{t}$, we express $r$ as a function of $\dot{f}$ and rewrite the angular momentum conservation in terms of the time derivatives of the angular coordinates only 
\begin{equation}
\left( G m_{\rm t} \right)^{2/3} m \dot{f}^{-1/3} + I_{\rm e} \dot{\psi} + I_{\rm c} \dot{\varphi} = J.
\label{amc_ang}
\end{equation}
Similarly, by applying Kepler III law, the second of equations~(\ref{eqs_motion}) can be rewritten  as:
\begin{equation}
\frac{1}{3} m r^{2} \ddot{f} = \frac{3}{8\pi} \frac{Gm_{\rm s}}{r^{3}} T \sin 2\alpha,  
\label{ddotf_eq}
\end{equation}
where we have restored the orbital radius after computing the time derivative of $m r^{2} \dot{f}$ to have the orbital moment of inertia in the l.h.s. of the equation. 

\begin{figure}
\centerline{
\includegraphics[height=8.0cm,width=8.0cm,angle=0,trim=1.5cm 1cm 1cm 0.5cm,clip]{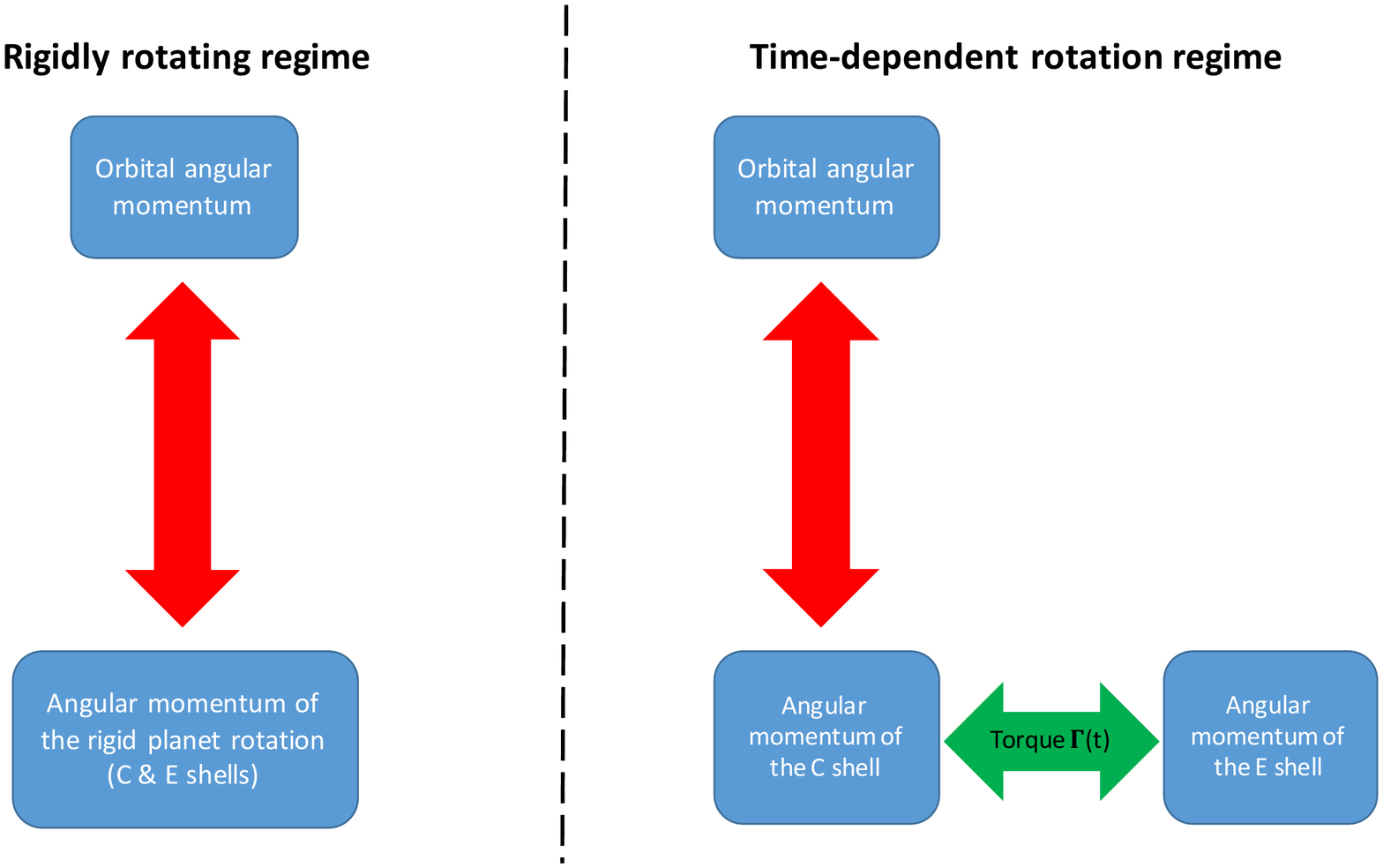}} 
\caption{Illustration of the angular momentum exchanges occurring in the two considered regimes. On the left, the  {\em rigidly rotating regime}, when the angular momentum is exchanged back and forth between the orbit and a rigidly rotating planet (red arrow) according to the simple pendulum equation~(\ref{pendulum_eq}). On the right, the  {\em time-dependent rotation regime} when the orbital angular momentum is exchanged between the orbit and the inner shell $C$ of the planet hosting its non-axisymmetric rigid core (red arrow), while the time-dependent torque $\Gamma (t)$ transfers angular momentum back and forth between the inner shell $C$ and the outer shell $E$ of the planet. The time-variable torque $\Gamma$ is produced by time-dependent Reynolds stresses in the envelope $E$ owing to vacillating or intermittent convection. }
\label{regimes}
\end{figure}

{The equations of motion of our dynamical system can be studied in two limiting regimes. The first occurs when the amplitude of the oscillations of the  rotation in the planetary interior is so small that it can be assumed to rotate rigidly (cf. Section~\ref{torsional_osc}), that is $\dot{\psi} \simeq \dot{\varphi}$. We call this the {\em rigidly rotating regime} (see Figure~\ref{regimes}, left panel). In this regime, all the net internal torques and, in particular, the torque between the $C$ and $E$ shells, can be considered negligible. Therefore, we can assume $\Gamma(t) =0$ in the equations of motion. On the other hand, the other regime occurs when the torque acting between  the $C$ and the $E$ shells is a periodic function of the time because of an intermittent or vacillating convection in the external shell $E$ (cf. Section~\ref{torsional_osc}).  The oscillation of $\Gamma$ periodically redistributes the angular momentum extracted from the orbital motion between the shells $C$ and $E$ (see Figure~\ref{regimes}, right panel). For simplicity, we shall consider the case when the amount of angular momentum exchanged between $C$ and $E$ in the course of an oscillation of $\Gamma (t)$ is much larger than the amount coming from the orbital motion and redistributed among $C$ and $E$ during the oscillation itself. We call this the {\em time-dependent rotation regime}. }

\subsubsection{The rigidly rotating regime}
\label{rigid_rot}
In this regime, the $C$ and $E$ shells rotate with the same angular velocity. Therefore, the angular momentum conservation equation becomes (cf. equation~\ref{amc}):
\begin{equation}
m r^{2} \dot{f} + I \dot{\varphi} = J,
\label{amc_rigid}
\end{equation}
where $I= I_{\rm c} + I_{\rm e}$ is the total moment of inertia of the planet. The torque $\Gamma(t)$  in the equations of motion vanishes  and we can obtain an equation for the angle $\alpha$ from the second and the third of equations (\ref{eqs_motion}) as
\begin{equation}
\ddot{\alpha} + \frac{1}{2} \omega_{\rm p}^{2} \sin 2\alpha = 0,
\label{pendulum_eq}
\end{equation}
that is, the equation of  motion of a simple pendulum making oscillations with a frequency $\omega_{\rm p}$ given by:
\begin{equation}
\omega_{\rm p}^{2} =  \frac{3}{4\pi} \frac{Gm_{\rm s}}{r^{3}} T \left( \frac{1}{I} - \frac{3}{mr^{2}} \right) \simeq \frac{3}{4\pi} n^{2} \left( \frac{T}{I} \right),
\label{omega_p}
\end{equation}
where $n$ is the mean orbital motion and we have made use of the Kepler III law and that $m_{\rm s} \simeq m_{\rm t}$ and $mr^{2} \gg I$. Note that $I$ appears in equation~(\ref{omega_p}) because the whole planet is rotating with the angular velocity $\dot{\varphi}$, thus the kinetic energy of rotation in the Lagrangian becomes $\frac{1}{2} I \dot{\varphi}^{2}$ giving the equation of motion for $\varphi$ -- the third of equations~(\ref{eqs_motion}) -- with $I_{\rm c}$ replaced by $I$. The moment of inertia of the orbit is of the order of $10^{3}$ times the moment of inertia of the planet, even in very close HJs, implying that $\omega_{\rm p}^2 >0$. 

Equation~(\ref{pendulum_eq}) admits the first integral:
\begin{equation}
\frac{1}{2} \dot{\alpha}^{2} + \frac{1}{2} \omega_{\rm p}^{2} \sin^{2} \alpha = \frac{1}{2} E_{0}^{2},
\label{first_integral}
\end{equation}
where $E_{0}$ is a constant of the motion that depends on the initial conditions. The equilibrium positions occur for $\alpha = \pm k \pi$, where $k$ is an integer, and correspond to $E_{0}=0$. The solutions of equation~(\ref{first_integral}) require $E_{0} \geq \omega_{\rm p} \sin \alpha$ because $\dot{\alpha}^{2} \geq 0$. If $E_{0} \leq \omega_{\rm p}$, the angle $\alpha$ librates around a position of equilibrium making oscillations with amplitude $\alpha_{0} = \arcsin(E_{0}/\omega_{\rm p})$ with $\dot{\alpha} = 0$ when $\alpha = \pm \alpha_{0}$. On the other hand, for $E_{0} > \omega_{\rm p}$ the angle $\alpha$ circulates, that is, it increases or decreases monotonously because $\dot{\alpha}$ never changes its sign. A change of the sign of $\dot{\alpha}$ would imply the quantity passing through zero which is not allowed by equation~(\ref{first_integral}) when $E_{0} > \omega_{\rm p}$. 

The period of libration is given by
\begin{equation}
P_{\rm libr} = \left( \frac{4}{\omega_{\rm p}} \right) K \left( \sin \alpha_{0} \right), 
\label{libr_per}
\end{equation}
where $K(\gamma)$ with $\gamma < 1$ is the complete elliptical integral of the first kind. The period diverges for $E_{0}/\omega_{\rm p} = \sin \alpha_{0} \rightarrow 1$ because $K({\gamma}) \rightarrow \infty$ when $\gamma \rightarrow 1$. On the other hand, the period of circulation is given by:
\begin{equation}
P_{\rm circ} = \left(\frac{4}{E_{0}} \right)  K \left( \frac{\omega_{\rm p}}{E_{0}} \right) 
\label{circ_per}
\end{equation}
that again diverges for $\omega_{\rm p}/E_{0} \rightarrow 1$ \citep[cf. the appendix in][]{Lanza20}. 

The variation in the orbital period $P$ associated with the transfer of angular momentum between the orbital motion and the spin of the planet can be computed from the variation of the true anomaly $f(t)$ because $P = 2\pi/\dot{f}$, giving:
\begin{equation}
\dot{P} = -\frac{1}{2\pi} P^{2} \ddot{f}.
\label{dotp_eq}
\end{equation}
Considering that $\alpha= f -\varphi$, applying the Kepler III law, and the conservation of the total angular momentum (equation~\ref{amc_rigid}), we find:
\begin{equation}
\ddot{f} = -\frac{I}{\frac{1}{3} m r^{2} - I} \ddot{\alpha} = \frac{I \omega_{p}^{2}}{2 \left(\frac{1}{3} m r^{2} - I \right)} \sin 2\alpha, 
\label{ddotf}
\end{equation}
where we substituted for $\ddot{\alpha}$ using equation~(\ref{pendulum_eq}). This equation is equivalent to 
equation~(\ref{ddotf_eq}) as can be shown by substituting the expression for $\omega_{\rm p}$ from equation~(\ref{omega_p}).  Equation~(\ref{dotp_eq})  becomes:
\begin{equation}
\dot{P} = -\frac{1}{4\pi} \frac{I\omega_{\rm p}^{2} P^{2}}{\left( \frac{1}{3}m r^{2} - I \right)} \sin 2 \alpha \simeq -\frac{3}{4\pi} \frac{I}{mr^{2}} \omega_{\rm p}^{2} P^{2} \sin 2\alpha, 
\label{dot_P}
\end{equation}
where we have considered that $mr^{2} \gg I$. From equation~(\ref{dot_P}), we see that the orbital period is modulated with a period $P_{\rm mod} = P_{\rm osc}/2$, where $P_{\rm osc}$ is the period of the oscillation of the angle $\alpha$ that is equal to $P_{\rm libr}$ in the case of libration or to $P_{\rm circ}$ in the case of circulation. 

The maximum relative orbital period variation is:
\begin{equation}
|\dot{P}|_{\max}/P \equiv \tau_{\rm P}^{-1} = \frac{3}{4\pi} \frac{I}{mr^{2}} \omega_{\rm p}^{2} P, 
\end{equation}
where we introduced the timescale for the orbital period variation $\tau_{\rm P}$. Substituting from equation~(\ref{omega_p}), we find:
\begin{equation}
|\dot{P}|_{\rm max}/P \simeq \frac{9}{4} \frac{T}{mr^{2}} \frac{1}{P}.
\end{equation}
For example, for a timescale $\tau_{\rm P} = 3$~Myr and a typical orbital period  $P=10^{5}$~s, we find $T/(mr^{2}) \sim 5 \times 10^{-10}$ or $T/I \sim 1.5 \times 10^{-6}$ assuming  that $I/(mr^{2}) \sim 3.3 \times 10^{-4}$. {Considering a Jupiter-like planet around a Sun-like star, equation~(\ref{quadrupole_deformation}) then gives an estimate of the core radius  $R_{\rm c} \sim 0.1$~$R_{\rm p}$, assuming $F=2$ and $k_{2} =0.36$, typical values of these parameters in the case of a Jupiter-like planet with a small core. }

The transit time variation $O-C$, produced by the periodic orbital period change, follows from equation~(\ref{oc_eq}) once $f(t)$  has been computed. By integrating equation~(\ref{ddotf}) with respect to the time, we find:
\begin{equation}
\Delta {f} (t) = -\frac{I}{\frac{1}{3} m r^{2} - I} \Delta {\alpha}(t) \simeq -\frac{3I}{m r^{2}} \Delta {\alpha} (t), 
\label{oc_computation}
\end{equation}
where any linear dependence of $\alpha(t)$ on the time has been subtracted because it corresponds to simply adjusting the constant orbital period of the reference orbit. 

In the case of libration, the excursion of the angle $\alpha$ is limited between $-\alpha_{0}$ and $\alpha_{0}$ with the limit $\alpha_{0} \rightarrow \pi/2$ corresponding to the limit $E_{0} \rightarrow \omega_{\rm p}$ giving an infinite libration period. Therefore, in the case of libration, the upper limit to the amplitude of the $O-C$ oscillation is:
\begin{equation}
\Delta (O-C)_{\rm libr} < \frac{1}{2} \left| \frac{IP}{\frac{1}{3}mr^{2} - I} \right| \simeq \frac{3}{2} \frac{I}{mr^{2}} P.
\label{maxoc_libr}
\end{equation}
In the case of circulation, we note that $\dot{\alpha}$ takes the same value when $\alpha$ varies by $\pi$ (cf. equation~\ref{first_integral}). In other words, if $P_{\rm circ}/2$ is the time taken by $\alpha$ to vary by $\pi$, we have $\dot{\alpha}(t + P_{\rm circ}/2) = \dot{\alpha(}t)$. Integrating this equation with respect to the time, we find that $\alpha(t+P_{\rm circ}/2)$ and $\alpha(t)$ must differ by a constant that, by definition, is equal to $\pi$. Generalizing this result and taking into account that $\alpha$ varies in a monotone way,  we have $\alpha(t + k P_{\rm circ}/2) = \alpha(t) \pm k\pi$, where $k$ is an arbitrary integer that is always positive or negative. Therefore, the maximum amplitude of the variation of $\alpha$ over one cycle of the orbital period modulation $P_{\rm mod} = P_{\rm circ}/2$, after subtracting the term that corresponds to a simple adjustment of the orbital period of the reference orbit, is $\pi$, giving again:
\begin{equation}
\Delta (O-C)_{\rm circ} \leq \frac{1}{2} \left| \frac{IP}{\frac{1}{3}mr^{2} - I} \right| \simeq \frac{3}{2} \frac{I}{mr^{2}} P.
\label{maxoc_circ}
\end{equation}
Considering that in the case of very close HJs, $P \sim 10^{5}$~s and $3I/(mr^{2}) \la 10^{-3}$, we have a maximum amplitude of the $O-C$ variations of $\sim 50$~s  both in the cases of libration or circulation of the angle $\alpha$. Therefore, oscillations of the $O-C$ having an  amplitude larger than $\sim 50$~s cannot be accounted for in the rigidly rotating regime. Note that in the case of the close stellar binary systems considered by \citet{Lanza20}, $3I/(mr^{2})$ ranges between $0.027$ and $0.12$, easily accounting for observed $O-C$ amplitudes that can reach several hours \citep[e.g.,][]{Muneeretal10}. 

In the case of a sinusoidal oscillation of the orbital period, the above results together with equation~(38) of \citet{Applegate92} allow us to evaluate the relative orbital period variation $\Delta P / P$ producing an $O-C$ amplitude of $A_{\rm O-C}$:
\begin{equation}
 \frac{\Delta P}{P}   = 2 \pi \frac{A_{\rm O-C}}{P_{\rm mod}} = 3 \pi \left( \frac{I}{mr^{2}} \right) \left( \frac{P}{P_{\rm mod}} \right). 
\end{equation}
Considering  $ P_{\rm mod} \sim 10^{9}$~s ($\approx 30$ yr), $P \sim 10^{5}$~s, and $I/(mr^{2}) \la 3.3 \times 10^{-4}$, we find a maximum $\Delta P / P \sim 3 \times 10^{-7}$. 

In addition to the limitation on the maximum  $O-C$, another difficulty of the rigidly rotating regime is the mechanism that excites the oscillations of the angle $\varphi$ because this requires a source of angular momentum external to the planet. Tides inside the planet tend to synchronize its rotation with the orbital motion damping the oscillations on a timescale of the order of $1-10$~Myr as we saw in Section~\ref{tides}. Therefore, the excitation of the oscillations of $\varphi$ must take place over at least that timescale. 

The simplest candidate to supply the required angular momentum is an impact with a body moving on a parabolic orbit and colliding with the planet. The amount of angular momentum required to excite the oscillations is $\Delta J_{\rm p} \leq  I \dot{\varphi}_{\rm max} \la I \omega_{\rm p}$, where we have applied the approximation $\dot{\alpha} \simeq -\dot{\varphi}$ because $|\ddot{f}| \ll |\ddot{\varphi}|$ and  equation~(\ref{first_integral}) to evaluate the maximum of $\dot{\alpha}$ as $\dot{\alpha}_{\rm max} \sim \omega_{\rm p} \approx \pi/P_{\rm mod}$. The angular momentum of a body moving on a parabolic orbit and colliding with the planet in a grazing impact  is $\Delta J_{\rm imp} \sim (2 G m_{\rm p} R_{\rm p})^{1/2} m_{\rm imp}$, where $m_{\rm imp}$ is the mass of the impactor.  Considering a planet with the mass, radius, and moment of inertia of Jupiter \citep{Nettelmannetal12}, and assuming $P_{\rm mod} = 30$ years, we find  $ m_{\rm imp} \sim 3.4 \times 10^{-4}$ Earth masses, implying a radius of $\sim 790$~km if its mean density is $10^{3}$~kg~m$^{-3}$. 

Although the possibility of an excitation of $\varphi$ oscillations by such a mechanism cannot be completely ruled out,  it seems rather unlikely that such an impact has occurred in the recent past ($1-10$~Myr) for a generic HJ system as  required to still see oscillations of significant amplitude, unless the system is very young so that many potential impactors are still available to collide with the planet. 

\subsubsection{The time-dependent rotation regime}
\label{inter_oscill}
{In this regime, the interior of the planet is not rotating rigidly, but the angular velocities of the $C$ and $E$ shells are oscillating with the dynamical coupling between them provided by the torque $\Gamma (t)$ produced by the time-dependent Reynolds stresses in the envelope $E$.  While the internal shell $C$ is rotating rigidly thanks to the strong Maxwell stresses produced by the magnetic field, the $E$ shell is in general rotating differentially (cf. Section~\ref{torsional_osc}). However, the time-dependent torque $\Gamma (t)$ that it applies to the inner shell $C$ produces a change of its mean angular velocity as measured by  $\dot{\psi}$. } For the sake of simplicity, we assume that the amount of angular momentum exchanged between the $C$ and $E$  shells during the oscillations of $\Gamma$ is much larger than that exchanged between the orbital motion and the rigid planetary core. 

By taking the difference of the fourth and the third of equations~(\ref{eqs_motion}) and neglecting the term containing $T$ because the corresponding angular momentum exchange is negligible in the above hypothesis, we find:
\begin{equation}
 \ddot{\beta} \simeq 	- \left( \frac{1}{I_{\rm c}} + \frac{1}{I_{\rm e}} \right) \Gamma(t), 
\label{beta_osc}
\end{equation}
where $\beta \equiv \psi - \varphi$. {When the function $\Gamma (t)$ is known, equation~(\ref{beta_osc}) can be integrated with respect to the time to find $\beta (t)$. In general, the specific function $\Gamma(t)$ depends on the  convection regime in the $E$ shell. Considering the numerical simulations by \citet{GastineWicht12} as a general guide, we can have simple sinusoidal oscillations in the vacillating regime of convection, more complex periodic functions in the intermittent regime, or erratic (non-periodic) oscillations in the chaotic regimes (cf. their Figure~3). Given our ignorance of the regime actually realized in hot Jupiters and having in mind an illustrative application of our model for the modulation of the orbital period, we consider a simple sinusoidal oscillation for $\Gamma(t)$, that leads to a sinusoidal oscillation of the angle $\beta(t)$. As we shall see below, this allows an analytic integration of equation~(\ref{ddotf_eq}) instead of requiring a numerical solution as in the general case. }

The solution of equation~(\ref{beta_osc}) can be coupled with the conservation of the internal angular momentum of the planet during the changes of its internal rotation that we write as
\begin{equation}
I_{\rm e} \Delta \psi(t) + I_{\rm c} \Delta \varphi(t) \simeq 0,
\end{equation}
introducing the variations of the angles $\varphi$ and $\psi$ with respect to the unperturbed regime of rigid rotation. In this way, we find 
\begin{equation}
\Delta \varphi(t) \simeq -\left( \frac{I_{\rm e}}{I} \right) \beta_{0} \cos \left( \omega_{\rm w} t + \varphi_{0} \right),
\end{equation}
where $\beta_{0}$ is the amplitude of the oscillation of the angle $\beta$, {$\omega_{\rm w}$ the pulsation  that is the same of that of the oscillations of the torque $\Gamma(t)$}, $I = I_{\rm c} + I_{\rm e}$, and $\varphi_{0}$ the initial phase of the oscillation. Since the variation of the true anomaly $f$ during a cycle of the orbital period modulation is much smaller than the variation of $\varphi$, we can write $\alpha(t) \simeq -\Delta \varphi(t)$, where we have subtracted the term giving a uniform variation of $\alpha(t)$ because it corresponds to a simple adjustment of the reference orbital period. Therefore, equation~(\ref{ddotf_eq}) becomes 
\begin{equation}
\ddot{f} =  F_{0} \sin \left[ A_{0} \cos (\omega_{\rm w} t +\varphi_{0}) \right], 
\label{ddotf_eq_fin}
\end{equation}
where 
\begin{equation}
F_{0} \equiv \frac{9}{8\pi} \left( \frac{Gm_{\rm s}}{r^{3}} \right) \left( \frac{T}{m r^{2}} \right)
\end{equation}
and
\begin{equation}
A_{0} \equiv 2 \left( \frac{I_{\rm e}}{I} \right) \beta_{0}.
\label{A0_beta0}
\end{equation}
The r.h.s. of equation~(\ref{ddotf_eq_fin}) is a periodic function of period $2\pi/\omega_{\rm w}$ that can be developed in a Fourier series by means of the Jacobi-Anger expansion \citep[see, e.g.,][\S~9.1.45]{AbramowitzStegun65} giving:
\begin{equation}
\ddot{f} =  2F_{0} \sum_{k=0}^{\infty} (-1)^{k} J_{2k+1} (A_{0}) \cos \left[ (2k+1) (\omega_{\rm w} t + \varphi_{0}) \right], 
\label{ddotf_fourier}
\end{equation}
where $J_{p} (z)$ is the Bessel function of the first kind of order $p$ and argument $z$. The series in the r.h.s. of equation~(\ref{ddotf_fourier}) can be integrated twice term by term giving the variation of the true anomaly as:
\begin{equation}
\Delta f(t) = - \frac{2F_{0}}{\omega_{\rm w}^{2}} \sum_{k=0}^{\infty} \frac{(-1)^{k} J_{2k+1}(A_{0})}{(2k+1)^{2}} \cos \left[ (2k+1) (\omega_{\rm w}t + \varphi_{0}) \right]. 
\label{deltaf_internal}
\end{equation}
This series has successive terms with alternating signs, therefore the error obtained by truncating the series at the order $q$ is given by the first neglected term, that is, the term of order $q+1$. 

The minimum amplitude $A_{0}$ can be evaluated by considering that the angular momentum is exchanged between the orbital motion and the planet core during an oscillation of the orbital period. In this way, the angular momentum conservation implies:
\begin{equation}
A_{0} \simeq  2 \pi \left( \frac{mr^{2}}{3I_{\rm c}} \right) \frac{(O-C)_{\rm max}}{P},  
\label{amp_A0}
\end{equation}
where we have applied equation~(\ref{oc_eq}) to compute the amplitude of $\Delta f$ from the maximum observed $O-C$ amplitude along a cycle of the orbital period. 

{In our model, a time-dependent differential rotation develops in the outer part of the planet. Its kinetic energy is dissipated by the turbulent eddy viscosity present in the same convective envelope and the power converted into heat is given by 
\begin{equation}
\dot{E}_{\rm kin} \simeq - \frac{1}{2} \int_{V_{\rm e}} \eta_{\rm t} s^2 \left( \frac{\partial \Omega}{\partial s} \right)^{2} \, dV,
\label{ekin_diss}
\end{equation}
where $\eta_{\rm t}$ is the turbulent dynamical viscosity in the convective shell $E$ of volume $V_{\rm e}$, $\Omega$ the angular velocity of rotation, and $s$ the cylindrical radius measured from the rotation axis \citep{LandauLifshitz59}. Equation~(\ref{ekin_diss}) assumes that the angular velocity depends only on the radial coordinate $s$ and the time as expected in the Taylor-Proudman regime and that the meridional flow is negligible. The turbulent viscosity is evaluated according to  \citet{Kichatinovetal94} taking into account the quenching of the standard mixing-length value due to the strong influence of the planetary rotation on the convective motions. 

To estimate the typical dissipated power expected in our model, let us consider a planet with the same internal structure and mass as Jupiter orbiting at a distance of 0.021~au from a star with the same mass of  the Sun with an orbital period $P =10^{5}$~s. The semiamplitude of the $O-C$ oscillation is assumed to be of 150~s with a modulation period of 25 yr. According to equation~(\ref{amp_A0}), this corresponds to $A_{0} = 51.3 $ radiants.  The  moments of inertia of the $C$ and $E$ shells are assumed to be $I_{\rm c}= 0.95 I$ and  $I_{\rm e} = 0.05 I$, respectively, where $I =  2.5  \times 10^{42}$~kg~m$^{2}$ is the moment of inertia of Jupiter. This corresponds to a base of the $E$ shell located at $\sim 0.9 R_{\rm J}$ as illustrated in Figure~\ref{inertia_model}. The angle $\beta_{0}$, computed from $A_{0}$ by means of equation~(\ref{A0_beta0}), is 513 rad, the semiamplitude of the relative variation of the rotation of the $C$ shell is $1.15 \times 10^{-3}$, while that of the $E$ shell is $0.022$ along the 25 yr of the modulation. 

We estimate the maximum amplitude of the radial differential rotation as $\partial \Omega / \partial s \simeq \omega_{\rm w} \beta_0/\delta_{\rm sh}$ by considering a typical shear length scale $\delta_{\rm sh} = 0.03 R_{\rm p}$ as suggested by \citet{HeimpelAurnou12} in the case of Jupiter, while the total radial extension of the layer over which the integration in equation~(\ref{ekin_diss}) is performed is taken of $0.1 R_{\rm p}$. In this way, we find a maximum dissipated power of $8.3 \times 10^{17}$~W that is larger than the power radiated by Jupiter of $ \approx 3.5 \times 10^{17}$~W.  However, assuming that a small fraction of the stellar insolation goes into powering the internal convection \citep{GuillotHavel11,YadavThorngren17}, we have enough power to support the maximum dissipation. Specifically, considering a Sun-like star and a planet with the radius of Jupiter separated by $0.021$~au, we have an insolation of $\sim 4.7 \times 10^{22}$~W, so that less than $\sim 0.002$ percent of the stellar luminosity intercepted by the planet is sufficient to account for the dissipation. \citet{ThorngrenFortney18} estimate that an insolation fraction between 0.2 and 2.5 percent is required to account for the inflated radii of hot Jupiters with masses greater than 0.5 Jupiter masses, which provides more than enough power to support the dissipation of the kinetic energy of the internal shear in our model. }

\section{Applications}

\subsection{Very hot Jupiters}
\label{result_vhjs}
We consider the list of very hot Jupiters in Table~1 of \citet{Patraetal20} because they are the most favourable targets to look for orbital decay and add two recently discovered hot Jupiters with very short orbital periods, that is, NGTS-6 \citep{Vinesetal19} and NGTS-10 \citep{McCormaketal19}. In Table~\ref{our_tabulation}, we list, from the left to the right, the name of the planetary system, the mass and the radius of the planet, the orbital period, the orbit semimajor axis,  the mass and the radius of the host star, and the maximum value of the $O-C$  computed in the case of the rigidly rotating regime (cf. Section~\ref{rigid_rot}). The planet and stellar parameters are taken from Table~1 of \citet{Patraetal20}, except for the two added NGTS systems for which they were extracted from the above discovery papers, respectively. 

\begin{figure}
\centerline{
\includegraphics[height=10.5cm,width=7.5cm,angle=90]{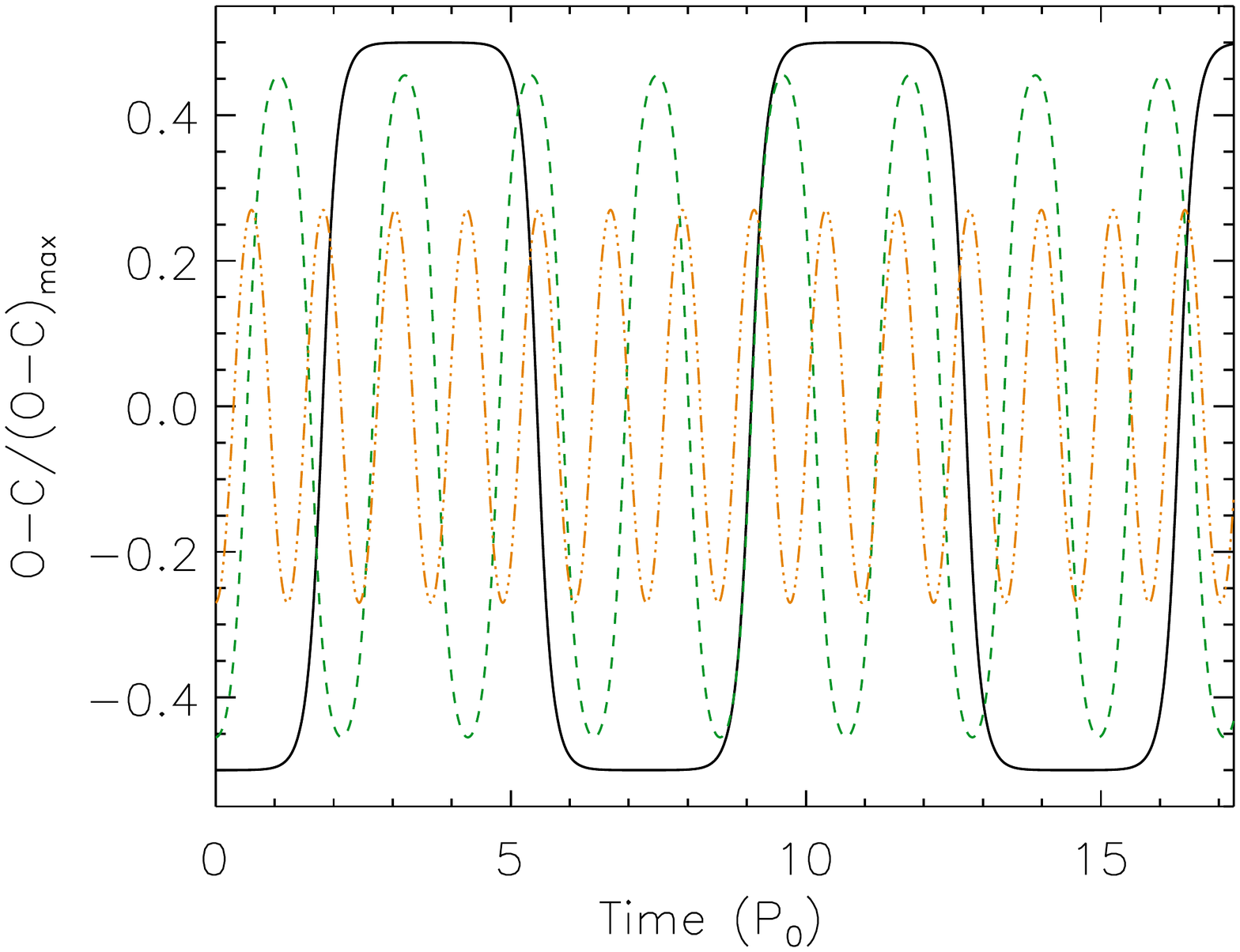}} 
\caption{Normalized $O-C$ in three cases of libration computed by means of equations~(\ref{pendulum_eq}) and ~(\ref{oc_computation}) vs. the time measured in units of the period of the oscillations in the linear regime, i.e., $P_{0} = 2\pi/\omega_{\rm p}$ (cf. Section~\ref{rigid_rot}). The normalization is to $(O-C)_{\rm max}$ as given by the upper limit in equation~(\ref{maxoc_libr}). Different linestyles and colours refer to different values of the limit angle $\alpha_{0}$: black solid line: $\sin \alpha_{0} = 1-10^{-8}$; green dashed  line: $\sin \alpha_{0} = 0.99$; and orange dash-three-dotted line: $\sin \alpha_{0} = 0.75$. }
\label{libration_fig}
\end{figure}

In Figure~\ref{libration_fig}, we plot the $O-C$ variations computed by numerically integrating equation~(\ref{pendulum_eq}) in the case of libration of the angle $\alpha$ for three values of the limit angle $\alpha_{0}$, that is, $\sin \alpha_{0} = 1-10^{-8}$, $0.99$, and $0.75$. The first is  close to the limit case ($\sin \alpha_{0} \rightarrow 1$) corresponding to the upper limit amplitude $(O-C)_{\rm max}$ and an infinite period because the complete elliptic integral in equation~(\ref{libr_per}) diverges (cf. Section~\ref{rigid_rot}). In this case, the $O-C$ curve is approaching a square wave with the variation of the orbital period concentrated within short time intervals around $P_{\rm libr}/2$ and its multiples. The period of the modulation is significantly longer than $P_{0} = 2\pi/\omega_{\rm p}$ because we approach the limit where the elliptical integral in equation~(\ref{libr_per}) diverges. On the other hand, the other two cases with a smaller $\sin \alpha_{0}$ correspond to oscillations of the $O-C$ with  amplitudes significantly smaller than the upper limit and periods closer to $P_{0}$, that is the period of the oscillations in the linear regime when $\sin 2\alpha \sim 2 \alpha$. 

\begin{figure}
\centerline{
\includegraphics[height=10.5cm,width=7.5cm,angle=90]{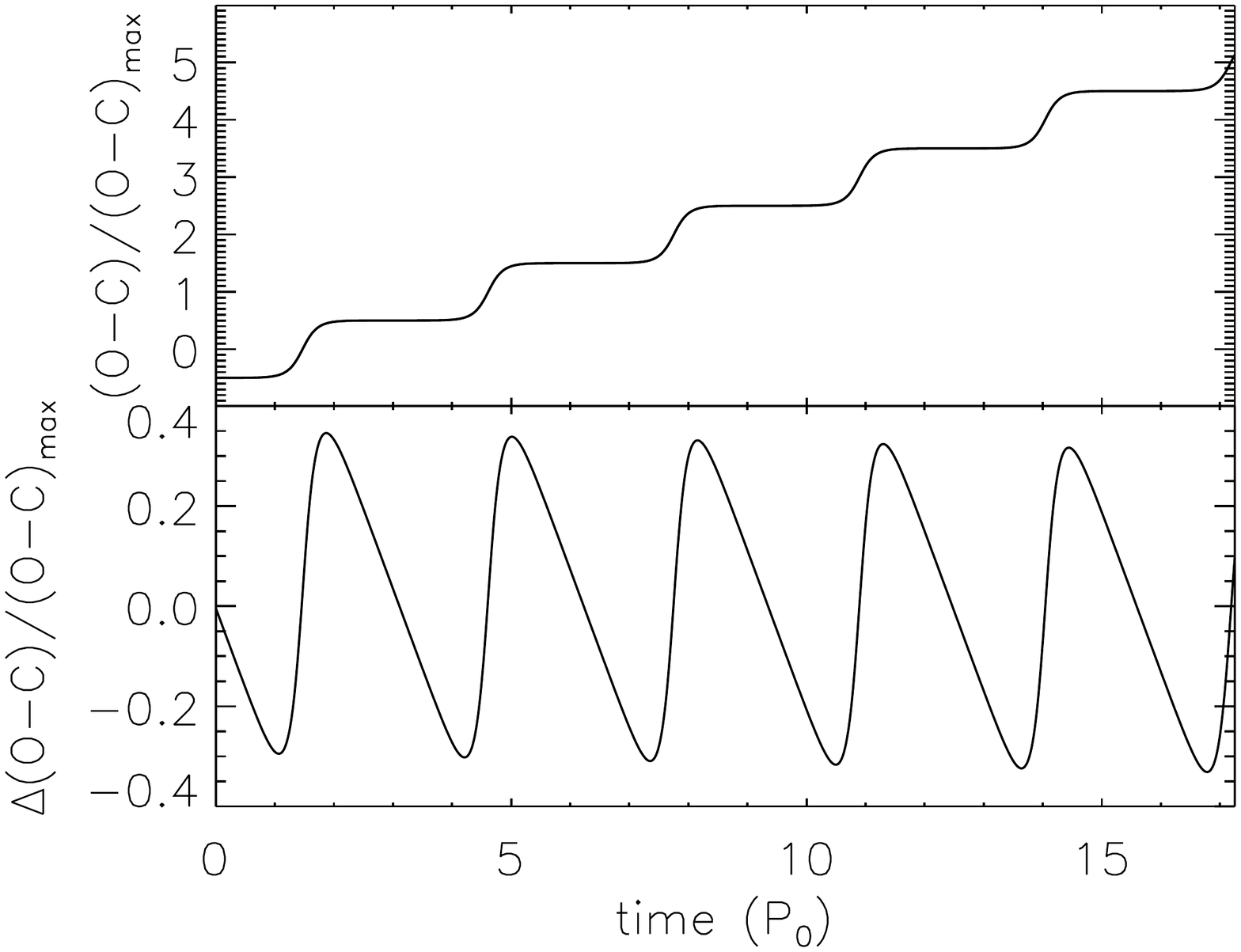}} 
\caption{Upper panel: Normalized $O-C$ in a case of circulation computed by means of equations~(\ref{pendulum_eq}) and~(\ref{oc_computation}) vs. the time measured in units of the period of the oscillations  in the linear regime, i.e., $P_{0} = 2\pi/\omega_{\rm p}$ (cf. Section~\ref{rigid_rot}). The normalization is to $(O-C)_{\rm max}$ as given by the maximum in equation~(\ref{maxoc_circ}). Lower panel: Same as the upper panel after subtracting the linear component of the $O-C$ variation  that corresponds to an adjustment of the reference orbital period. }
\label{circulation_fig}
\end{figure}

In the upper panel of Figure~\ref{circulation_fig}, we plot the normalized $O-C$ vs. the time obtained in a case of circulation of the angle $\alpha$, while in the lower panel we plot the $O-C$ after removing the linear variation in the upper panel that corresponds to adjusting the reference orbital period to isolate the modulation (cf. Section~\ref{rigid_rot}). The initial conditions have been selected in order to produce an $O-C$ amplitude that is $\sim 0.8$ of the maximum amplitude. This gives an oscillation period that is remarkably longer than $P_{0}$. Cases with a smaller residual amplitude give a period closer to $P_{0}$ as in the case of libration. 

The maximum amplitudes of the $O-C$ oscillations for the systems in Table~\ref{our_tabulation} were computed by means of equations~(\ref{maxoc_libr}) and~(\ref{maxoc_circ}) that give the same value both in the case of libration and circulation of the angle $\alpha$. Such a value depends on the ratio $I/(mr^{2}) \simeq h_{\rm p} (R_{\rm p}/r)^{2}$, where we have used the normalized moment of inertia of the planet $h_{\rm p}= I/(m_{\rm p} R_{\rm p}^{2})$ (cf. Section~\ref{tides}) and the approximation $m \simeq m_{\rm p}$. Since $h_{\rm p} \sim 0.26$ depends on the internal structure of the planet and is likely to be more or less constant among hot Jupiters \citep{Guetal03}, the maximum $O-C$ depends essentially on the ratio $R_{\rm p}/r$ and the orbital period $P$ becoming larger for larger values of those parameters. 

We see that the rigidly rotating regime is not capable of accounting for the amplitude of the $O-C$ observed in WASP-12 because it is a factor of $\sim 3-4$ times larger than the maximum amplitude predicted by the model \citep{Yeeetal20,Patraetal20}. Therefore, we shall consider the time-dependent rotation regime for this system in Section~\ref{results_wasp12}. On the other hand, the marginally significant $O-C$ found in the case of WASP-19 \citep{Patraetal20} is within the limit predicted by the rigidly rotating regime. However, the problem of exciting the oscillations of the whole planet remains open (cf. Section~\ref{rigid_rot}). 

\begin{table*}
\begin{tabular}{lccccccc}
\hline
               Name & $m_{\rm p}$ & $R_{\rm p}$ & $P$ & $a$ & $m_{\rm s}$ & $R_{\rm s}$ & $(O-C)_{\rm max}$ \\
                  & (m$_{\rm J})$ & (R$_{\rm J})$ & (d) & (au) & (M$_{\odot})$ & (R$_{\odot})$ & (s) \\ 
\hline
     WASP-18 & 11.40 &  1.20 &  0.94 &  0.021 &  1.46 &  1.29 &  23.86 \\
     KELT-16 &  2.75 &  1.42 &  0.97 &  0.020 &  1.21 &  1.36 &  37.27 \\
    WASP-103 &  1.51 &  1.62 &  0.93 &  0.020 &  1.21 &  1.42 &  49.91 \\
     WASP-12 &  1.47 &  1.90 &  1.09 &  0.023 &  1.43 &  1.66 &  57.42 \\
     HATS-18 &  1.98 &  1.34 &  0.84 &  0.018 &  1.04 &  1.02 &  39.00 \\
     WASP-19 &  1.14 &  1.41 &  0.79 &  0.016 &  0.94 &  1.02 &  47.13 \\
  OGLE-TR-56 &  1.39 &  1.36 &  1.21 &  0.024 &  1.23 &  1.36 &  31.99 \\
    HAT-P-23 &  2.09 &  1.37 &  1.21 &  0.023 &  1.13 &  1.20 &  33.79 \\
     WASP-72 &  1.55 &  1.27 &  2.22 &  0.037 &  1.39 &  1.98 &  20.89 \\
     WASP-43 &  2.03 &  1.04 &  0.81 &  0.015 &  0.72 &  0.67 &  29.94 \\
    WASP-114 &  1.77 &  1.34 &  1.55 &  0.029 &  1.29 &  1.43 &  27.31 \\
    WASP-122 &  1.28 &  1.74 &  1.71 &  0.030 &  1.24 &  1.52 &  46.08 \\
          NGTS-6 &  1.34 &  1.33 &  0.88 &  0.017 &  0.77 &  0.75 &  44.40 \\
           NGTS-10 &  2.16 &  1.21 &  0.77 &  0.014 &  0.70 &  0.70 &  42.98 \\
\hline
\end{tabular}
\caption{Parameters of the sample of very hot Jupiters considered in our model application together with the maximum $O-C$ amplitude obtained in the rigidly rotating  regime (see the text). The mass of Jupiter and its radius are indicated as $\rm m_{\rm J}$ and $\rm R_{\rm J}$, while the mass of the Sun and its radius as $\rm M_{\odot}$ and $\rm R_{\odot}$, respectively. }
\label{our_tabulation}
\end{table*}

\subsection{WASP-12}
\label{results_wasp12}
The $O-C$ diagram of WASP-12 has been modelled with a parabola corresponding to a constant orbital period decrease \citep{Yeeetal20,Patraetal20}. We assume that the observed $O-C$ variation over about $9.5$ years is part of a longer-term modulation with a period of $\approx 25$~years produced by a cyclic exchange of angular momentum between the orbital motion and the core of the planet.  {In the time-dependent rotation regime, such an exchange is produced by the modulation of the rotation of the core shell $C$ owing to the time-dependent Reynolds stresses at the interface with the envelope $E$ as discussed in Section~\ref{torsional_osc}}. 

\begin{figure}
\centerline{
\includegraphics[height=10.5cm,width=7.5cm,angle=90]{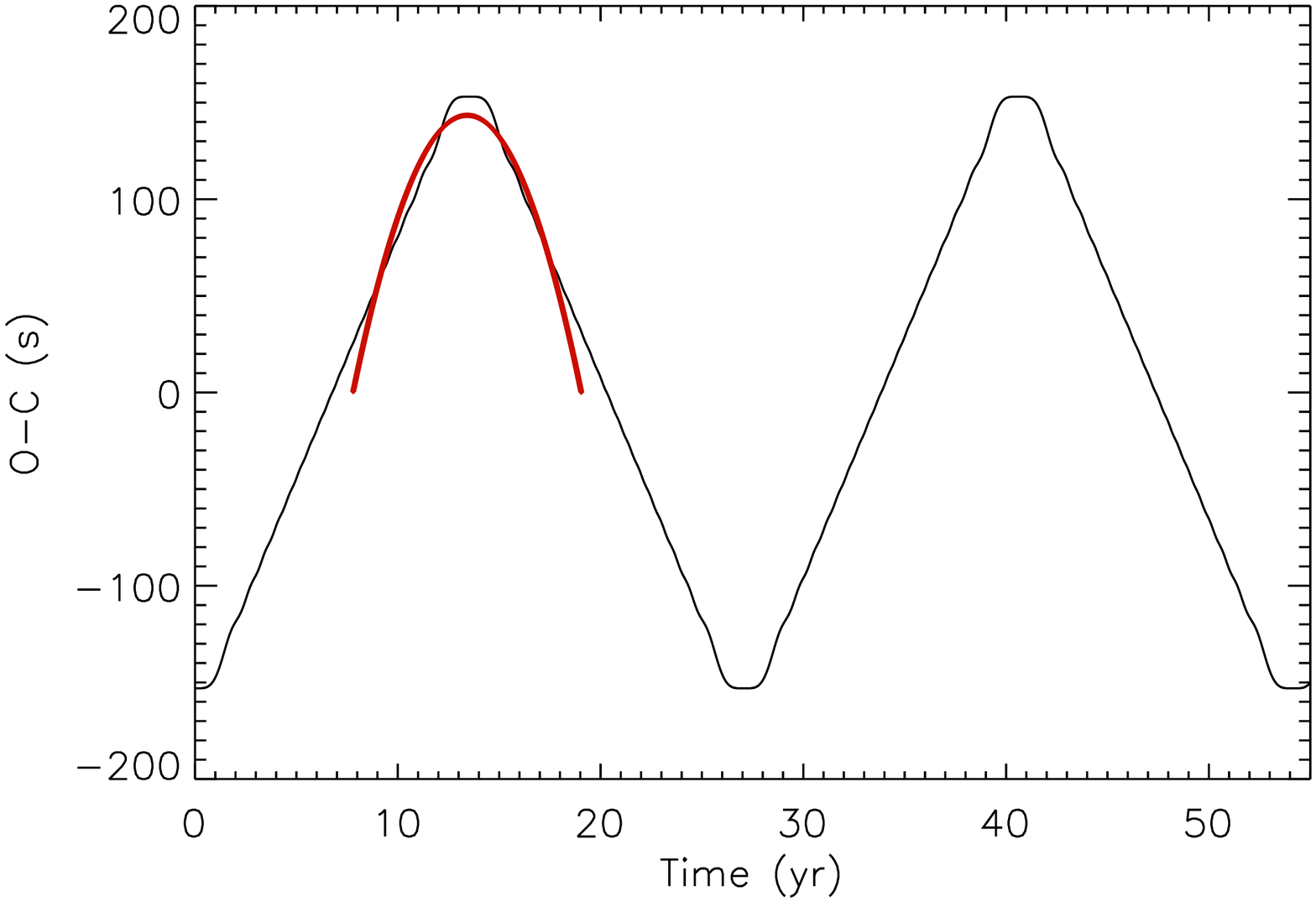}} 
\caption{$O-C$ predicted in the time-dependent rotation regime vs. the time in the case of WASP-12 assuming the parameters listed in Table~\ref{wasp12_param} (black line) {and an initial phase $\varphi_{0}=\pi$ in equation~(\ref{deltaf_internal}).} The red line shows a parabola that approximates the observed $O-C$ variation within the errors on a timescale of $\sim 10$~years (see the text). }
\label{wasp12_model}
\end{figure}

In Figure~\ref{wasp12_model} we show an illustrative model computed with the parameters listed in Table~\ref{wasp12_param}. {We assume that the moments of inertia of the $C$ and $E$ shells are $I_{\rm c} = 0.95\,I$ and $I_{\rm e} =0.05\,I$, where $I$ is the moment of inertia of the planet, respectively.  We consider the simplest case of a sinusoidal oscillation of the torque $\Gamma(t)$ acting on the core shell $C$ of the planet as expected in the case of vacillating convection because this regime admits an analytic solution of the problem. The $O-C$ modulation plotted in Figure~\ref{wasp12_model} is obtained from equation~(\ref{oc_eq}), where $\Delta f$ is computed by means of the series in equation~(\ref{deltaf_internal}) truncated at order 50, which gives a truncation error $< 2\times 10^{-19}$~s. } The observed parabolic $O-C$ variation has an amplitude of $\sim 150$~s over a time interval of $\sim 10$~year \citep[cf. Fig. 19 of][]{Patraetal20} as indicated by the red line in Fig.~\ref{wasp12_model}. We do not attempt a direct fitting of the observations because such a fit would be of limited value given the limited time extension of the available data, their typical errors of the order of $\ga 30$~s in the single $O-C$ measurements, and the number of free parameters of our model that makes the best fit parameters not unique. The deviation of the model $O-C$ variations from the parabola is well within the observational errors, thus the difference between the two is not detectable with the present data. Future space-borne photometry may provide individual $O-C$'s with errors of $5-10$~s, thus allowing to discriminate between the two representations of the orbital period change. 

{We adopted an oscillation period $P_{\rm mod}$ of 27.1~years for the torque $\Gamma(t)$, and adjusted the permanent gravitational quadrupole moment $T$ of the core in order to reproduce a semiamplitude of the $O-C$ modulation of approximately 150~s to account for the available observations. Nevertheless, different combinations of $P_{\rm mod}$ and $T$ can be found that reproduce equally well the observed $O-C$ variation owing to its limited time extension.  }

{The quadrupole moment of the core $T$ is related to the core radius $R_{\rm c}$ through the Love number $k_{2}$,  measuring its tidal deformation, and the factor $F$ parametrizing the effect of the fluid envelope on the deformation of the solid core (cf. Section~\ref{core_potential}). Adopting the parameters in Table~\ref{wasp12_param}, and considering equation~(\ref{quadrupole_deformation}) with $k_{2}=0.36$ and $F=2$, appropriate for  a giant planet with a small core, we find $R_{\rm c} = 0.07$~$R_{\rm p}$, assuming that the star-planet separation has not changed since the core solidified. } This $R_{\rm c}$ is actually an upper limit for the radius of the core because the star-planet separation could have been smaller than the present value when the core solidified soon after its formation. After that initial phase,  stellar tides could have pushed the young planet outwards if the stellar rotation period was initially shorter than the orbital period with the subsequent orbital evolution accounting for the presently observed separation \citep{BolmontMathis16}. 

{The amplitude  $A_{0} = 56.6$~rad in our model; the relative semiamplitude of the oscillation of the angular momentum of the rigid core $C$ is $1.6 \times 10^{-3}$, while that of the envelope shell $E$ is $0.031$.  These values are comparable with the relative amplitudes predicted by the model of \citet{HeimpelAurnou12} or the numerical simulations of \citet{GastineWicht12} as discussed in Section~\ref{torsional_osc}. 

The power dissipated by the action of the turbulent convection on the differential rotation in the interior of the planet is given by equation~(\ref{ekin_diss}). It can be estimated by considering a simple dimensional scaling of the shear and the dynamic turbulent viscosity with the radius and the mass of the planet which gives $\dot{E}_{\rm kin} \propto M_{\rm p} R_{\rm p}^{2}$. Considering the similarity of $A_{0}$ and of the modulation period in the example  in Section~\ref{inter_oscill} and the planet parameters in Table~\ref{wasp12_param}, the maximum power dissipated inside WASP-12b turns out to be $4.4 \times 10^{18}$~W.  
Given that the stellar insolation is $\sim 6 \times 10^{23}$~W, less than $ 10^{-5}$ of the insolation is enough to supply the maximum power dissipated by the internal shear during the modulation of the orbital period. }

WASP-12 is presently the only system showing a significant deviation from a constant orbital period, therefore, we do not provide an application of our model to other very hot Jupiters, although it can be  computed from the theory in Section~\ref{inter_oscill}. The lack of significant $O-C$ detections in the case of the other very hot Jupiters can be interpreted assuming that oscillations of their internal rotation of a sufficiently large amplitude as to produce measurable orbital period variations do not occur, at least over the timescale of one decade covered by current observations \citep[see][]{Patraetal20}. In this case, only the rigidly rotating planet regime is viable to produce an orbital period modulation in the framework of our assumptions, thus the lack of a significant $O-C$ may indicate that no oscillations of the whole planet rotation have been excited in those systems or that their cores lack a permanent quadrupole moment. 

\section{Discussion and conclusions}
We have introduced a model for the orbital period modulation in systems with close-by giant planets based on a coupling between the spin of the planet core and the orbital motion. The coupling is produced by a permanent non-axisymmetric quadrupole moment in the planet core that produces a torque on the orbital motion allowing a cyclic exchange of angular momentum between the planet rotation and the orbit. This requires that the planet core be solid, while no orbital period change can be produced in the case of a fluid core. An alternative to a solid core to maintain a permanent quadrupole moment could be an internal stationary magnetic field with an intensity of the order of at least $10^{3}$~G as proposed by \citet{Lanza20} in the case of late-type stars. Even an axisymmetric field can produce a non-axisymmetric gravitational quadrupole moment, provided that it is inclined to the planet spin axis. However, given our present ignorance about the dynamo operating in hot Jupiters, we do not further explore this possibility. 

The angular momentum exchanged between the orbit and the core spin is then redistributed inside the planet. {We investigated two regimes of internal angular momentum transport corresponding to a rigidly rotating planet and to a planet with an internal time-dependent rotation, respectively. When the planet is rotating rigidly, there is an upper limit to the amplitude of the $O-C$ modulation of $\approx 50$~s, while when the internal rotation is time dependent, we can account for larger $O-C$ amplitudes. Moreover,  the mechanism in the former  case needs to be excited by large impacts producing a temporary deviation from a regime of synchronous rotation for the planet, otherwise enforced by the stellar tides on timescales as short as $1-10$~Myr. 

In the latter case, a vacillating or intermittent convection in the outer convective envelope of the planet can produce a cyclic modulation of the Reynolds stresses at the interface with the inner part of the planet,  thus producing a cyclic oscillation of its angular velocity.  For example, in WASP-12b, a variation of the rotation of the inner part with a semiamplitude of $\sim 0.16$ percent is sufficient to produce an orbital period modulation with a semiamplitude of the $O-C$ of $\sim 150$~s along a cycle of $\sim 25$~yr, capable of accounting for the observations. 
In this case, the exchange of angular momentum produces a variation of the angular velocity of the outer envelope with a semiamplitude of $\sim 3$~percent along the modulation cycle. It could be detectable if the variation in the surface zonal flows leads to a longitude shift of the hot spot produced by the stellar irradiation in the planetary atmosphere. 

Recent investigations suggest that the surface zonal flows of Jupiter and Saturn could extend down to the level where the transition from the molecular to the metallic state of hydrogen and helium produces a remarkable increase of the conductivity coupling the flow to the internal magnetic fields of the planets \citep{Christensenetal20}. Therefore, such flows could reveal variations of the angular momentum of the outer convective shells of hot Jupiters, rather than being a probe only of the circulation in the outermost layers of their atmospheres. In such a case, we may expect a variation in the longitude of the hot spots observed close to the planet occultations in transiting hot Jupiters. Nevertheless, other processes, such as the interaction with the planetary magnetic field or inhomogeneous clouds, may alter the longitude of the hot spots requiring a careful analysis to disentangle the expected small rotational variation from other possible effects \citep{Dangetal18}. }

We applied our model to a sample of very hot Jupiters to evaluate the maximum $O-C$ amplitude in the case of the rigidly rotating planet regime. This regime cannot account for the amplitude of the $O-C$ variations observed in WASP-12, therefore we proposed an illustrative application of our model in the time-dependent internal rotation regime for this system. Our model predicts that the observed negative curvature of its $O-C$ diagram will sooner or later be reversed because it is part of a modulation whose period depends on the {time-dependent Reynolds stresses produced by convection in the outer shell of the planet}, transferring angular momentum back and forth between that shell and the interior of the planet. However, different combinations of the model parameters are possible, thus our guess of the duration of the orbital period modulation cycle should be regarded as a lower limit. 

WASP-12 is presently the only system showing a significant deviation from a constant-period ephemeris. The lack of similar detections in the case of other very hot Jupiters could be an indication of the lack of a solid core with a permanent non-axisymmetric quadrupole deformation; {or the indication of an almost stationary convection in their outer shells;} or could be due to the short time intervals sampled by the observations. In other words, other systems with a significant orbital period change may emerge in the future because the sensitivity to the period change increases quadratically with the time span of the observations. WASP-19 is a candidate system, but a longer time baseline is needed to confirm or disprove its change. 

A non-axisymmetric solid core is a necessary condition for the operation of our model, but it is not sufficient because we need to excite oscillations  of the planet rotation as a whole in the case of a rigidly rotating interior or {in the form of a time-dependent radial differential rotation in the time-dependent regime.} If these additional conditions are not satisfied, our mechanism cannot work and no period modulation is expected. Therefore, only a relatively small fraction of very hot Jupiters may satisfy all the conditions for the operation of our mechanism. WASP-12 could be one of those systems thanks to its large inflated radius that suggests an internal heat source  that could power a strong dynamo action and {a non-stationary convection regime} in its interior. In this hot Jupiter, the strong stellar insolation may ultimately be responsible for both the large radius inflation and {the vacillating or intermittent convection regime} as required by our model.  
\begin{table}
\begin{tabular}{lc}
\hline
Planet radius $R_{\rm p}$ (R$_{\rm J}$) & 1.90\\
Planet mass $m_{\rm p}$ (M$_{\rm J}$) & 1.47\\
Orbit semimajor axis $r$ (au) & 0.02344\\
Star radius $R_{\rm s}$ (R$_{\odot}$) & 1.657\\
Star mass $m_{\rm s}$ (M$_{\odot}$) & 1.434\\
Orbital period $P$ (d) & 1.0914 \\
Amplitude $A_{0}$ (rad) & 56.6\\
Period of the orbital modulation (yr) & 27.07 \\ 
Cylindrical radius  of the $C$-$E$ interface $s_{0}$ ($R_{\rm p}$) & 0.90\\
Moment of inertia of the core $I_{\rm c}/I$ (with $I=0.26\, m_{\rm p} R_{\rm p}^{2}$) & 0.95 \\
Core quadrupole moment $T/(mr^2)$ & $2.1\times 10^{-9}$\\
\hline
\end{tabular}
\caption{Parameters of the model of WASP-12 orbital period modulation in the time-dependent rotation regime (see the text). Note that all the parameters come from previous measurements or are assumed in order to have an $O-C$ semiamplitude of $\sim 150$~s, except for the value of $T/(mr^{2})$ that results from our model. }
\label{wasp12_param}
\end{table}
\section*{Data availability}
The data underlying the applications of the model presented in this article are available from the references cited in Sections 4.1 and 4.2.
\section*{Acknowledgements}
The author is grateful to an anonymous referee for several comments that greatly helped him in improving the model and its presentation. He also acknowledges support by INAF/Frontiera through the "Progetti Premiali" funding scheme of the Italian Ministry of Education, University, and Research. 
%
%




\bsp	
\label{lastpage}
\end{document}